\documentclass[10pt, conference, compsocconf]{IEEEtran}
\usepackage{cite}
\usepackage{makecell}
\usepackage{caption}
\usepackage{amsmath,amssymb,amsfonts}
\usepackage{algorithmic}
\usepackage{graphicx}
\usepackage{textcomp}
\usepackage{xcolor}
\usepackage{booktabs}
\usepackage{subfigure}
\usepackage{subcaption}
\usepackage{algorithm}
\usepackage{pifont}
\usepackage{flushend}

\usepackage{amsthm}

\newtheorem{theorem}{Theorem}            

\theoremstyle{definition}                

\newcommand{\toolname}[0]{}
\renewcommand{\toolname}[0]{ECORE }
\def\BibTeX{{\rm B\kern-.05em{\sc i\kern-.025em b}\kern-.08em
    T\kern-.1667em\lower.7ex\hbox{E}\kern-.125emX}}

\newcounter{insight}
\newenvironment{insight}{
  \refstepcounter{insight}
  \par\noindent\textbf{Insight \#\theinsight.} 
}{\par}
\def\BibTeX{{\rm B\kern-.05em{\sc i\kern-.025em b}\kern-.08em
    T\kern-.1667em\lower.7ex\hbox{E}\kern-.125emX}}
\begin{document}

\title{ECORE: Energy-Conscious Optimized Routing for Deep Learning Models at the Edge}

\author{
\IEEEauthorblockN{
Daghash K. Alqahtani$^{1}$,
Maria A. Rodriguez$^{2}$,
Muhammad Aamir Cheema$^{3}$,
Hamid Rezatofighi$^{4}$,
Adel N. Toosi$^{1}$
}
\IEEEauthorblockA{
$^{1}$Distributed Systems and Network Applications (DisNet) Laboratory, The University of Melbourne, Australia\\
$^{2}$School of Computing and Information Systems, The University of Melbourne, Australia\\
$^{3}$Department of Software Systems and Cybersecurity, Monash University, Australia\\
$^{4}$Department of Data Science and AI, Monash University, Australia\\
\{daghash.alqahtani@student, maria.read, adel.toosi\}@unimelb.edu.au,
\{aamir.cheema, hamid.rezatofighi\}@monash.edu
}
}

 \maketitle

\begin{abstract}
Edge computing enables data processing closer to the source, significantly reducing latency, an essential requirement for real-time vision-based analytics such as object detection in surveillance and smart city environments. However, these tasks place substantial demands on resource-constrained edge devices, making the joint optimization of energy consumption and detection accuracy critical. To address this challenge, we propose \toolname, a framework that integrates multiple dynamic routing strategies, including a novel estimation-based techniques and an innovative greedy selection algorithm, to direct image processing requests to the most suitable edge device–model pair. \toolname dynamically balances energy efficiency and detection performance based on object characteristics. We evaluate our framework through extensive experiments on real-world datasets, comparing against widely used baseline techniques. The evaluation leverages established object detection models (YOLO, SSD, EfficientDet) and diverse edge platforms, including Jetson Orin Nano, Raspberry Pi 4 and 5, and TPU accelerators. Results demonstrate that our proposed context-aware routing strategies can reduce energy consumption and latency by 35\% and 49\%, respectively, while incurring only a 2\% loss in detection accuracy compared to accuracy-centric methods.
\end{abstract}


\section{Introduction}
\label{sec:intro}


Object detection, a core task in computer vision analytics, has been widely adopted across domains such as agriculture, sports analytics, transportation, healthcare, and smart cities. Recent advancements in machine learning and artificial intelligence (AI) have significantly enhanced the performance of object detection models. As a result, deploying these models on edge platforms such as Raspberry Pi and NVIDIA Jetson has become an effective strategy for edge computing, thereby reducing latency, preserving privacy, and enabling scalable, on-device intelligence. In particular, deep learning-based models such as \textit{You Only Look Once (YOLO)}~\cite{Redmon2016}, \textit{Single shot multibox detector (SSD)}~\cite{Liu2016} and \textit{EfficientDet}~\cite{Tan2020} 
come in multiple variants to accommodate different computational budgets. 
For instance, the YOLOv8~\cite{ultralytics} family includes \textit{nano}, \textit{small}, \textit{medium}, and \textit{large} versions each reflecting trade-offs between model size, speed, and accuracy.

The dynamic nature of object detection in real-world environments presents both a challenge and an opportunity. Visual scenes vary widely in complexity, from sparsely populated streets at night to dense daytime urban intersections, leading to fluctuating computational demands for object detection. In such contexts, the heterogeneity of edge resources, including both hardware and detection models, offers an opportunity for intelligent allocation of requests to device–model pairs. 

Our recent findings show that detection models exhibit markedly different energy, latency, and accuracy profiles when deployed on diverse edge devices~\cite{alqahtani2024}. By leveraging this hardware–model diversity, systems can achieve substantial energy savings without compromising detection accuracy or latency. Energy efficiency is particularly critical on edge devices operating under strict power constraints, such as those relying on batteries or local energy harvesting methods like solar panels. Yet in safety-critical applications like surveillance and traffic monitoring, accuracy remains non-negotiable. This creates the central challenge addressed in this study: ``\textit{how can we minimize energy usage without degrading detection quality?}''

Our proposed solution lies in adaptive, context-aware routing strategies that dynamically select the optimal device–model pair based on scene complexity. Simpler scenes allow for lightweight processing, while complex environments require more accurate--and consequently more energy-intensive--models. To address this, we propose the \toolname framework, which introduces a set of dynamic and intelligent routing algorithms to redirect an image processing task to the most appropriate device-model pair. The goal is to reduce energy consumption while maintaining high detection accuracy. Note that this routing mechanism should be lightweight, so it does not offset the energy savings, since any added overhead increases the overall image processing pipeline cost.

ECORE's routers use lightweight estimation techniques during the pre-processing stage to infer the number of objects in each incoming image as an indicator of scene complexity. To the best of our knowledge, this is the first work to address the routing problem through the lens of scene complexity. Based on this estimate, the router selects and forwards the request to an optimal device-model pair. The main contributions of this work are as follows:
\begin{enumerate}
\item We propose a novel system architecture, \toolname, that optimizes the routing of image processing tasks across heterogeneous edge devices running diverse object detection models, coordinated by a central gateway.

\item We introduce three object count estimation techniques to enable intelligent routing in \toolname: (i) a lightweight edge-detection–based method using Canny, (ii) an SSD-based lightweight detection model deployed at the gateway, and (iii) a low-overhead approach that exploits object detection results from previous inference rounds.

\item We conduct an extensive experimental evaluation on a real-world testbed, comparing our approaches against an Oracle Router and multiple baselines using realistic datasets including COCO and a surveillance video stream under diverse scenarios.



\end{enumerate}


\section{Motivation}
\label{sec:motivation}
\begin{figure}[t]

    \subfigure[]{
        \includegraphics[width=0.46\columnwidth]{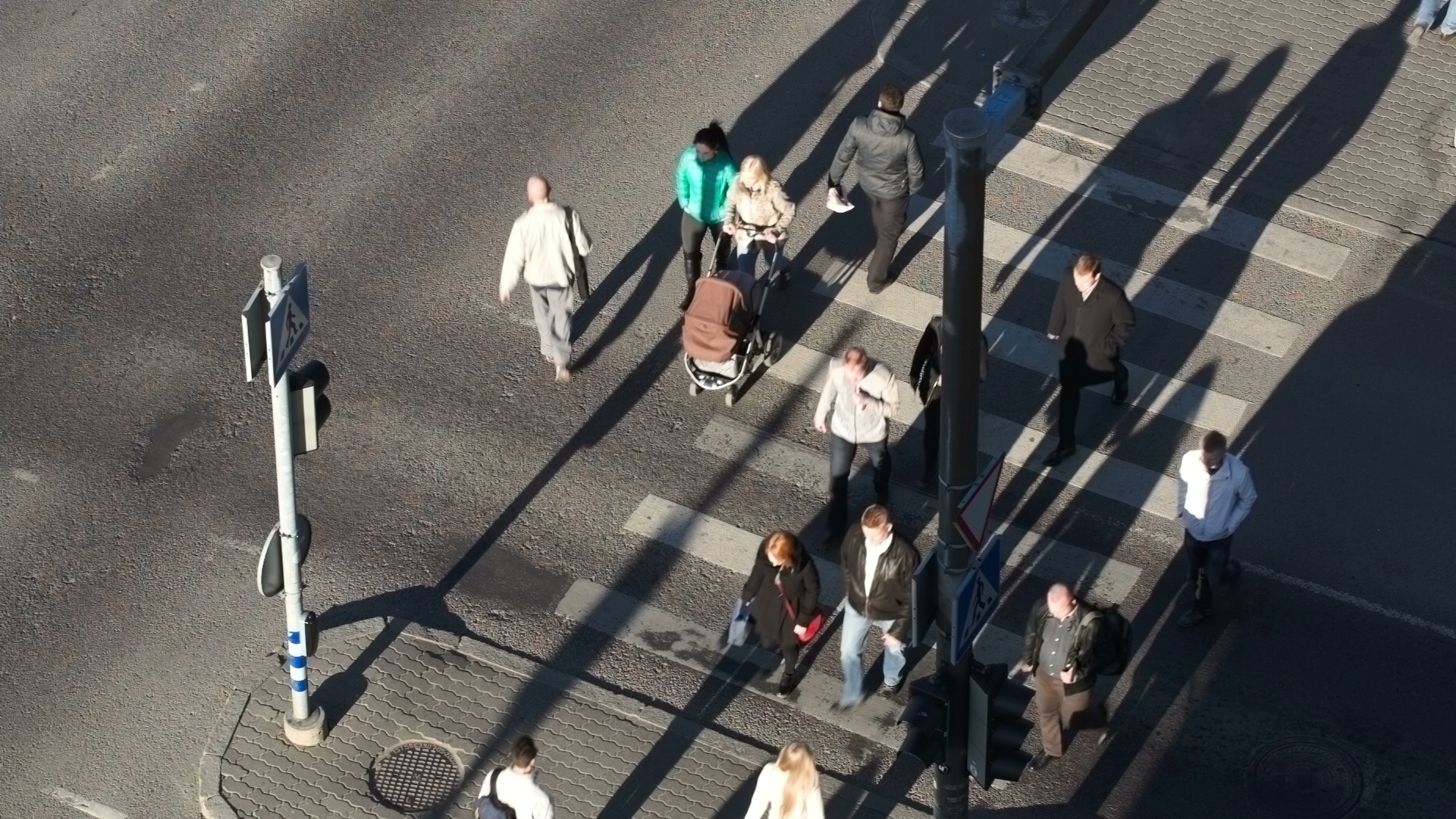} 
    }
  \subfigure[]{
        \includegraphics[width=0.46\columnwidth]{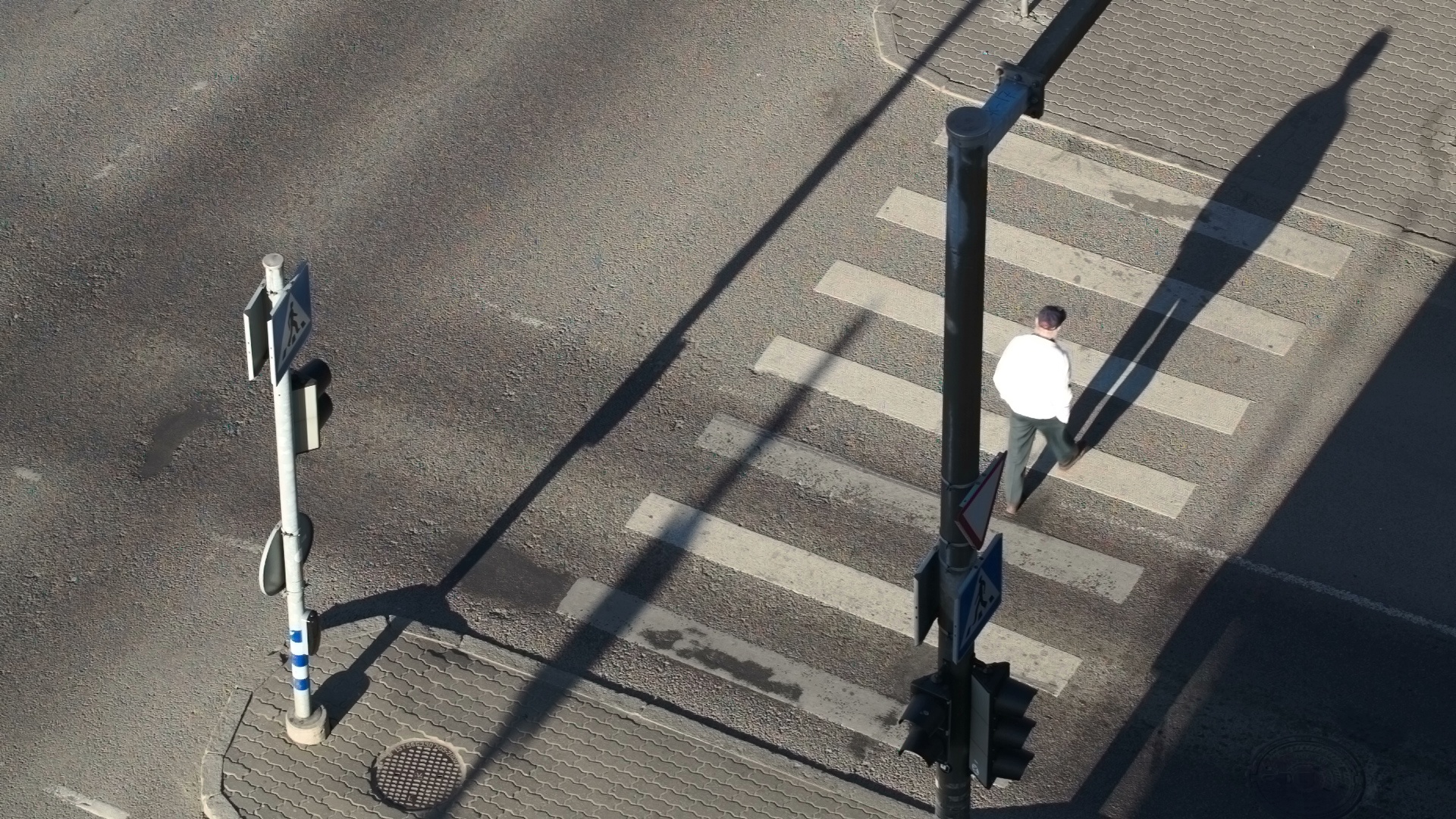}
    }
\vspace{-0.3cm}
\caption{Contrasting pedestrians scenarios: (a) High density scene; (b) Low density scene.}
\label{fig:scene_examples}
\vspace{-0.2cm}
\end{figure}

 Fig.~\ref{fig:scene_examples} illustrates two contrasting scenes captured by an urban surveillance camera at an intersection. The first image depicts numerous pedestrians crossing the intersection, while the second image shows the same location when it is sparsely populated. In the former high density situation, accurate object detection becomes more critical and at the same time more computationally demanding, especially in safety sensitive applications such as surveillance, traffic monitoring, and crowd management. 
 In contrast, the latter depicts a 
 low-activity scenario in which the computational effort required to achieve reasonable accuracy is reduced, and the risk of critical misdetections is substantially lower. 
 Thus, energy savings can be achieved by using lighter detection models, particularly when edge devices operate under battery or power constraints.
This motivates the design of our proposed routing algorithms, which seeks to optimize energy consumption without compromising accuracy where it matters most.

\begin{figure}[htbp]

\subfigure[Energy Consumption per Request.]{
        \includegraphics[width=0.46\columnwidth]{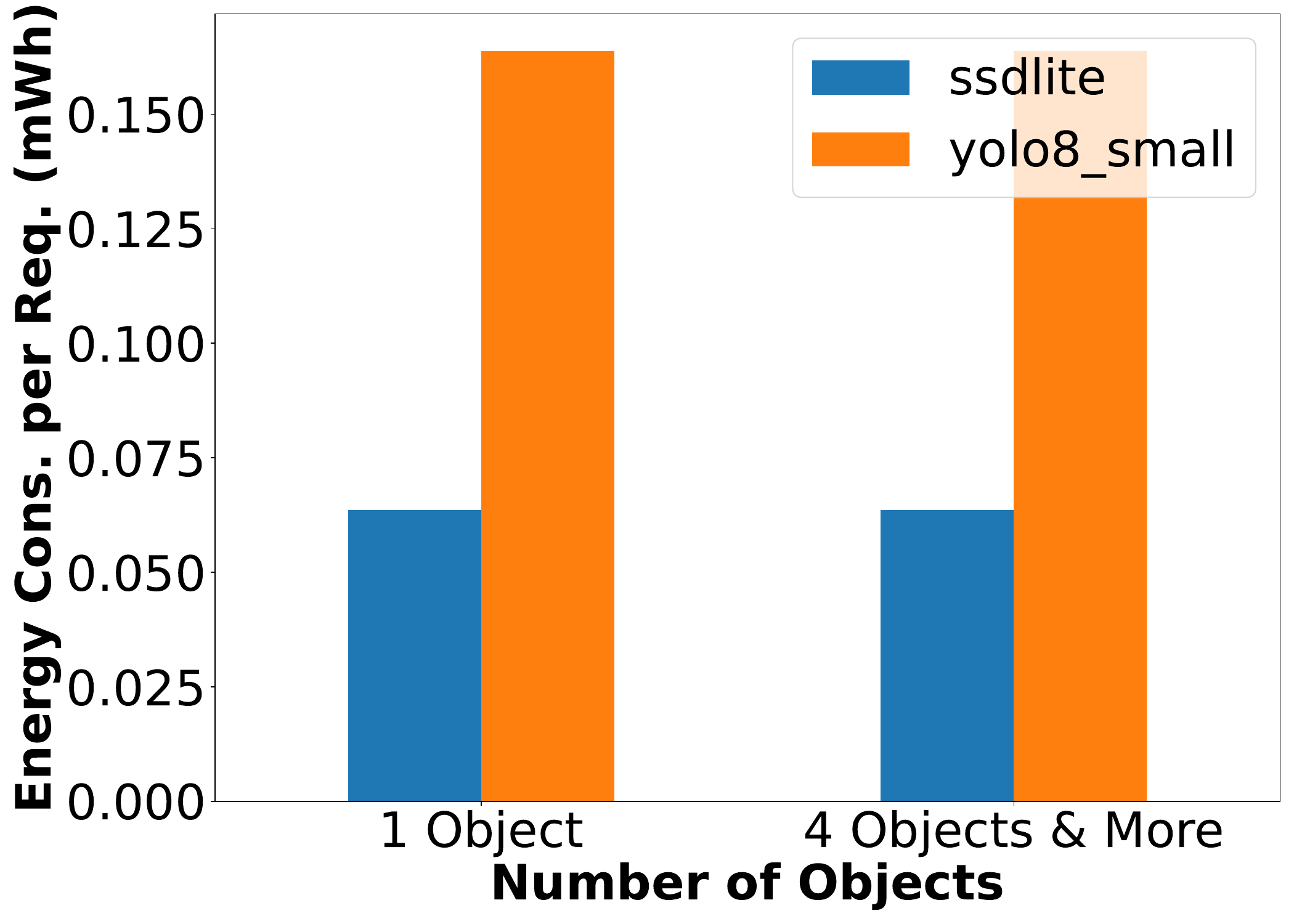}
    }
    \subfigure[Accuracy (mAP)]{
        \includegraphics[width=0.46\columnwidth]{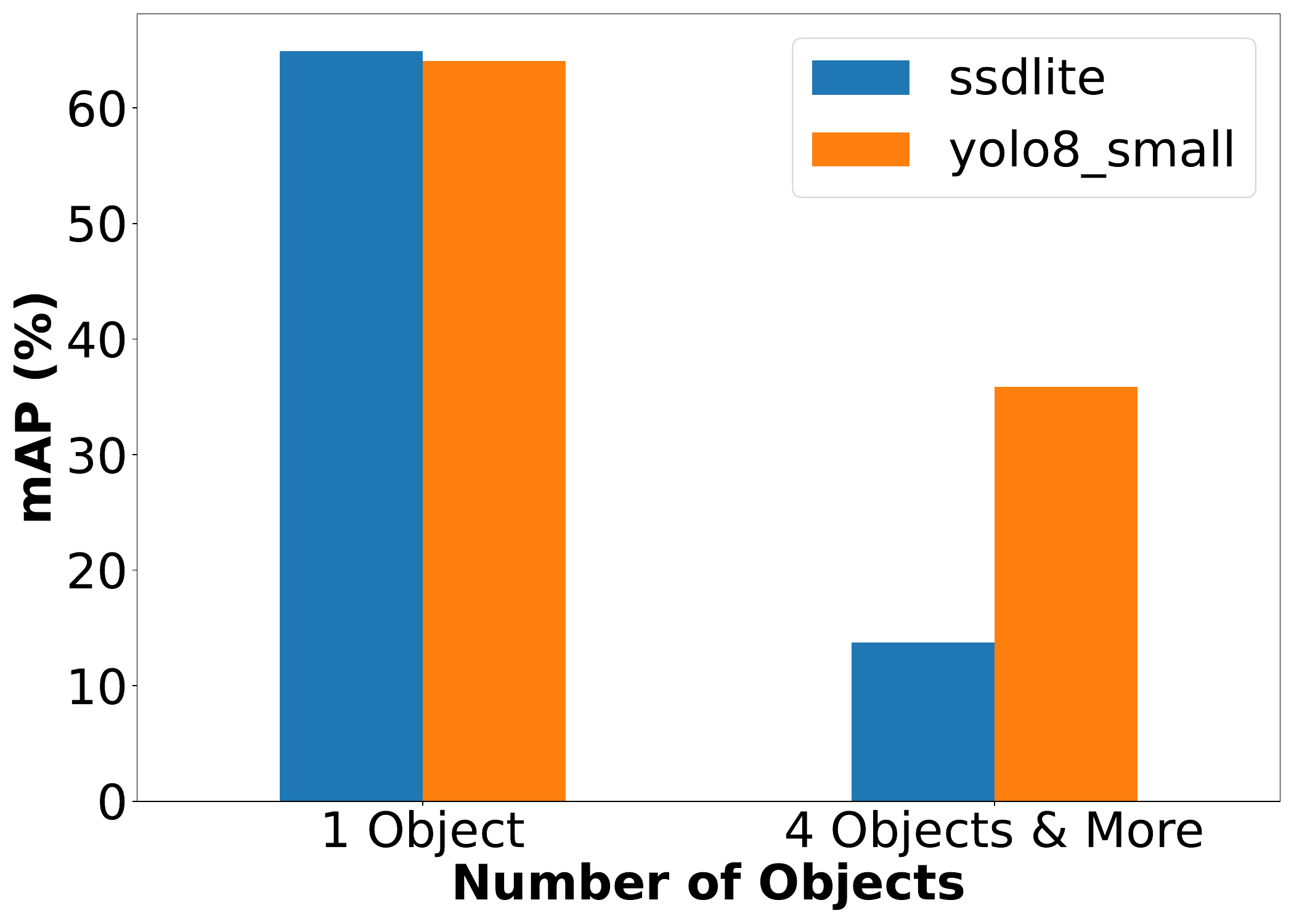}
    }
\caption{Energy Consumption and Accuracy for different group of images across object detection models.}
\label{fig:energy and mAP}
\vspace{-0.2cm}
\end{figure}

To verify our hypothesis, we conducted a preliminary experiment using the COCO validation dataset, which contains 5,000 images. We categorized the images into two groups based on the number of objects in the image, as determined by the ground truth: one group with a single object per image and another with four or more. We then calculated the mAP and energy consumption per image during inference. As shown in Fig.~\ref{fig:energy and mAP}(b), SSD Lite and YOLOv8 small exhibit similar accuracy for single-object images, while YOLOv8 small significantly outperforms SSD Lite on images with four or more objects, achieving nearly twice better mAP. In terms of energy usage, SSD Lite consumes a consistent amount across both groups, which is approximately 50\% lower than YOLOv8 small. These results suggest that in less complex scenarios, two different object detection models can achieve similar accuracy, while significant energy savings is possible by using more efficient models such as SSD Lite without compromising accuracy.

\section{System Design and Implementation}
\label{sec:system design}
\begin{figure}[t]
\centerline{\includegraphics[width=0.50\textwidth]{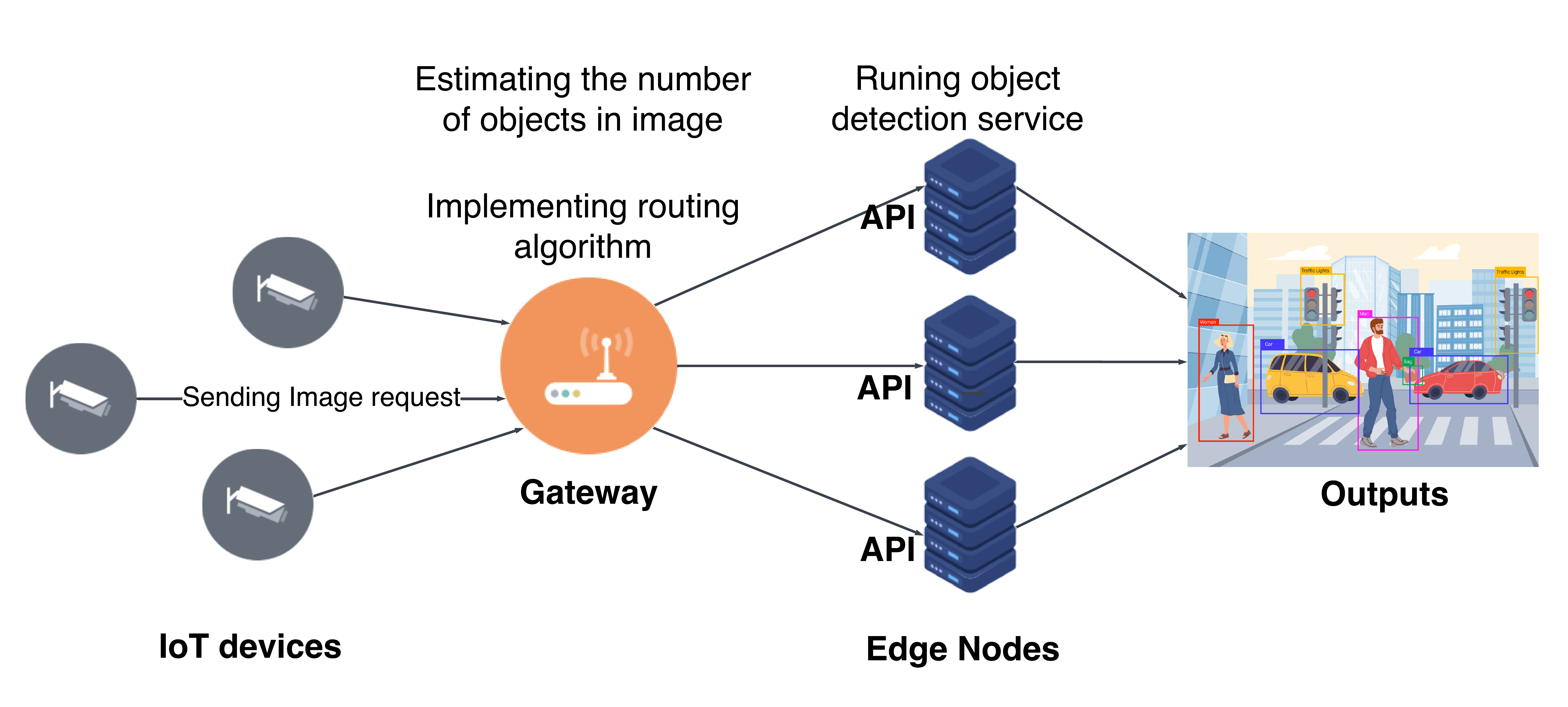}}
\caption{System architecture presentation.}
\label{system_arch}
\vspace{-0.2cm}
\end{figure}

This section describes the system architecture of \toolname, designed for energy-efficient and performance-aware routing of object detection tasks in an edge environment. Fig.~\ref{system_arch} illustrates the overall design. We assume a scenario where a pool of backend edge devices hosts a variety of device-model pairs. Each device runs a dedicated object detection model and exposes an inference API to process images or video frames captured via distributed cameras or embedded sensors. Cameras or embedded sensors transmit image data or video frames to a central gateway, which performs lightweight preprocessing and request coordination. This gateway-centric design reflects common smart-surveillance deployments (e.g., CCTV), decoupling cameras from device heterogeneity and centralizing monitoring and decision-making. The same workflow also applies to other pervasive edge settings where sensors generate inference requests and multiple heterogeneous edge nodes are available (e.g., smart retail, traffic monitoring, industrial inspection, and campus IoT). As shown in Fig.~\ref{system_arch}, the gateway applies one of the proposed object estimation techniques to infer the number of objects in each incoming image. This object-estimation component is modular, and alternative estimators can be substituted without changing the rest of the routing pipeline. The estimated count is passed to the routing algorithm (Algorithm~\ref{alg:routing}), which uses it to select the most appropriate device-model pair.

The routing algorithm begins by identifying the corresponding object count group based on the estimated number of objects, using predefined group rules defined by numeric ranges and labels. It then filters the profiling data to retain only the entries that match the identified group. From this subset, it calculates the maximum achievable mAP and derives a minimum acceptable mAP within a predefined tolerance margin. The data is then filtered again to retain only device-model pairs whose mAP falls within this target range. Among these, the pair with the lowest energy consumption is selected. The gateway forwards the request to the selected endpoint, which processes the image and returns the detected objects and their bounding boxes.

\subsection{Problem Formulation and Greedy Approach}

When running an object detection application in an edge environment, a key challenge lies in routing image requests to the most suitable device-model pair, taking into account trade-offs among accuracy, latency, and energy. Let:
\[
\mathcal{M} = \{(m_1, d_1), (m_2, d_2), \dots, (m_n, d_n)\}
\]
denote the set of available object detection device-model pairs, where 
$m_i$ represents a specific object detection model and $d_ i$ denotes the device on which it can be executed. Each pair \( (m_i, d_i) \in \mathcal{M} \) is profiled in advance, yielding key performance indicators: 1) \( \text{mAP}_i \): mean Average Precision; 2) \( t_i \): inference time (in milliseconds); 3) \( e_i \): energy consumption (in milliwatt-hour (mWh)); and 4)\( g_i \): object count group (e.g. 0, 1, 2, 3, etc).

Given an incoming image request accompanied by an estimated object count group \( G \), the routing objective selects the most energy-efficient device-model pair that satisfies application constraints for accuracy. This decision-making process can be formulated as a constrained optimization problem. First, the subset \( \mathcal{M}_G \subseteq \mathcal{M} \) is identified, containing only the pairs belonging to group \( G \). From this subset, the algorithm determines the maximum achievable accuracy.
\[
\text{mAP}_{\text{max}} = \max_{i \in \mathcal{M}_G} \text{mAP}_i
\]

To preserve flexibility in device-model selection and allow for more energy-efficient decision, we introduce a dynamic tolerance margin, \( \delta_{\text{mAP}} \).
This controlled relaxation enables the system to trade off a small reduction in mAP for gains in energy efficiency. The margin defines a feasible set \( \mathcal{F} \subseteq \mathcal{M}_G \), within which device-model pairs are considered acceptable for routing, and formally defined as:
\[
\mathcal{F} =\{ i \in \mathcal{M}_G \mid \text{mAP}_i \geq \text{mAP}_{\text{max}} - \delta_{\text{mAP}}\}
\]

The routing algorithm then selects the optimal pair by minimizing energy consumption:
\[
i^* = \arg\min_{i \in \mathcal{F}} e_i.
\]

\begin{algorithm}[t]
\caption{Routing Algorithm}
\begin{algorithmic}[1]
\REQUIRE
  number\_of\_objects,
  profiling\_data,
  $\delta$-mAP,
  group\_rules

\STATE \textbf{Determine} \texttt{group} by searching \texttt{group\_rules}:
\FORALL{\((\texttt{range},\texttt{label})\in\texttt{group\_rules}\)}
  \IF{\(\texttt{number\_of\_objects}\in\texttt{range}\)}
    \STATE \(\texttt{group} = \texttt{label}\)
    \STATE \textbf{break}
  \ENDIF
\ENDFOR
\STATE \textbf{Filter} \(\texttt{profiling\_data}\) to select only rows where 
\(\texttt{object\_count\_group} = \texttt{group}\)
\STATE \textbf{return} \texttt{group\_data}\
\STATE \textbf{Compute} \(\texttt{max\_mAP}\) = maximum \texttt{mAP} (\texttt{group\_data})
\STATE \textbf{Compute} \(\texttt{mAP\_min} =\texttt{max\_mAP} - \texttt{delta\_mAP}\)
\STATE \textbf{Filter} \(\texttt{group\_data}\) to retain only rows with 
\(\texttt{mAP} \ge \texttt{mAP\_min}\)
\STATE \textbf{return} \(\texttt{refined\_data}\)
\STATE \textbf{Select} the row in \(\texttt{refined\_data}\) that has \textbf{the lowest \(\texttt{energy}\)}
\STATE \textbf{return} the device–model
\end{algorithmic}
\label{alg:routing}
\end{algorithm}

This greedy algorithm, as detailed in Algorithm~\ref{alg:routing}, provides a principled balance between performance and efficiency, enabling the system to route requests to energy-efficient configurations with only marginal degradation in accuracy. It is particularly suitable for real-time applications deployed in resource constrained environments, where the router must make fast decisions under uncertain and dynamic workloads.

\subsection{Optimality of the Routing Algorithm}
While the routing problem proposed in the previous section involves trade-offs among multiple objectives, applying threshold-based filtering prior to selection reduces the decision to a one-dimensional optimization. This simplification not only lowers computational complexity but also guarantees optimality under the  model assumptions. As a result, the algorithm remains robust, fast, and well-suited for deployment in resource-constrained, real-time environments. The structure of the problem ensures that the greedy solution is indeed optimal. The following theorem formally proves the optimality of the greedy routing algorithm presented in Algorithm~\ref{alg:routing}.

\begin{theorem}[Optimality of the Routing Algorithm]
Let $\mathcal{M}_G$ be the subset of profiled device-model pairs matching the estimated object count group $G$, and let $\delta_{\text{mAP}}$ be a fixed accuracy tolerance. Then, the greedy selection of the device-model pair with the lowest energy consumption from the filtered feasible set 
\[
\mathcal{F} = \left\{ i \in \mathcal{M}_G \mid \text{mAP}_i \geq \text{mAP}_{\max} - \delta_{\text{mAP}} \right\}
\]
yields an optimal solution to the routing problem under the specified constraints.
\end{theorem}

\begin{proof}
The objective is to minimize energy consumption:
\[
\min_{i \in \mathcal{F}} e_i
\]
where the feasible set $\mathcal{F}$ includes only device-model pairs in $\mathcal{M}_G$ whose accuracy exceeds the threshold $\text{mAP}_{\max} - \delta_{\text{mAP}}$.

Since all the constraints including accuracy and group relevance are applied during the filtering step, the remaining selection is unconstrained and reduces to choosing the element with minimum $e_i$. 

As the energy consumption values $e_i$ are known from prior uncorrelated profiling, the greedy algorithm that selects
\[
i^* = \arg\min_{i \in \mathcal{F}} e_i
\]
is guaranteed to return the globally optimal solution to the constrained minimization problem. There are no interdependencies or temporal dynamics that would invalidate this selection.

\end{proof}

\subsection{Proposed object count estimation approaches}

To avoid negating energy savings, the routing algorithm must remain lightweight, especially its object-counting component. We therefore design and evaluate three approaches for estimating the number of objects in incoming image requests and directing them to the most suitable edge node.

\noindent\textbf{--Edge Detection (ED):} This approach leverages edge detection which is a fundamental technique in image processing and computer vision used to identify areas of significant intensity or color change, which often correspond to object boundaries. The primary goal is to capture essential structural information. We employ the widely used Canny edge detection algorithm~\cite{canny1986}, to estimate the number of objects in each image. The resulting object count is then fed into the proposed routing algorithm (Algorithm~\ref{alg:routing}), which selects the optimal device-model pair based on the expected energy–accuracy trade-off governed by $\delta_{\text{mAP}}$. This method is considered lightweight and computationally efficient, though it may yield coarse or approximate estimations.

\noindent\textbf{--SSD-Based Front-End (SF):} In this approach, we deploy a very lightweight and efficient Single Shot MultiBox Detector (SSD) object detection model directly at the gateway. The model processes incoming images to identify and count objects, and the resulting count is passed to the routing algorithm, which determines the appropriate edge node. Compared to the edge detection method, this approach is expected to provide more accurate object count estimates, albeit at a higher computational cost at the gateway.

 \noindent\textbf{--Output-Based (OB):} This approach leverages temporal continuity in image streams, particularly video, by assuming that consecutive frames contain a similar number of objects. Instead of estimating the object count for every image (frame), the system reuses the count obtained from the output of the previous image (frame), as detected by the last selected device-model pair. The same configuration is retained until a change in object count is observed.
    The routing algorithm begins with a default estimate (e.g., zero) and forwards the first image to an edge node. Once the detection results are returned, the router extracts the actual number of objects and uses it for subsequent routing decisions. If the object count remains stable, the system continues using the same device-model pair; if a change is detected, the routing decision is updated accordingly. This adaptive strategy avoids per-image (frame) estimation, reducing computational overhead and improving energy efficiency. 

\section{Performance Evaluation}
\label{sec:performance evaluation}
\subsection{Experimental Setup}
\subsubsection{Dataset}

\begin{figure}[t]
\centerline{\includegraphics[width=0.8\columnwidth]{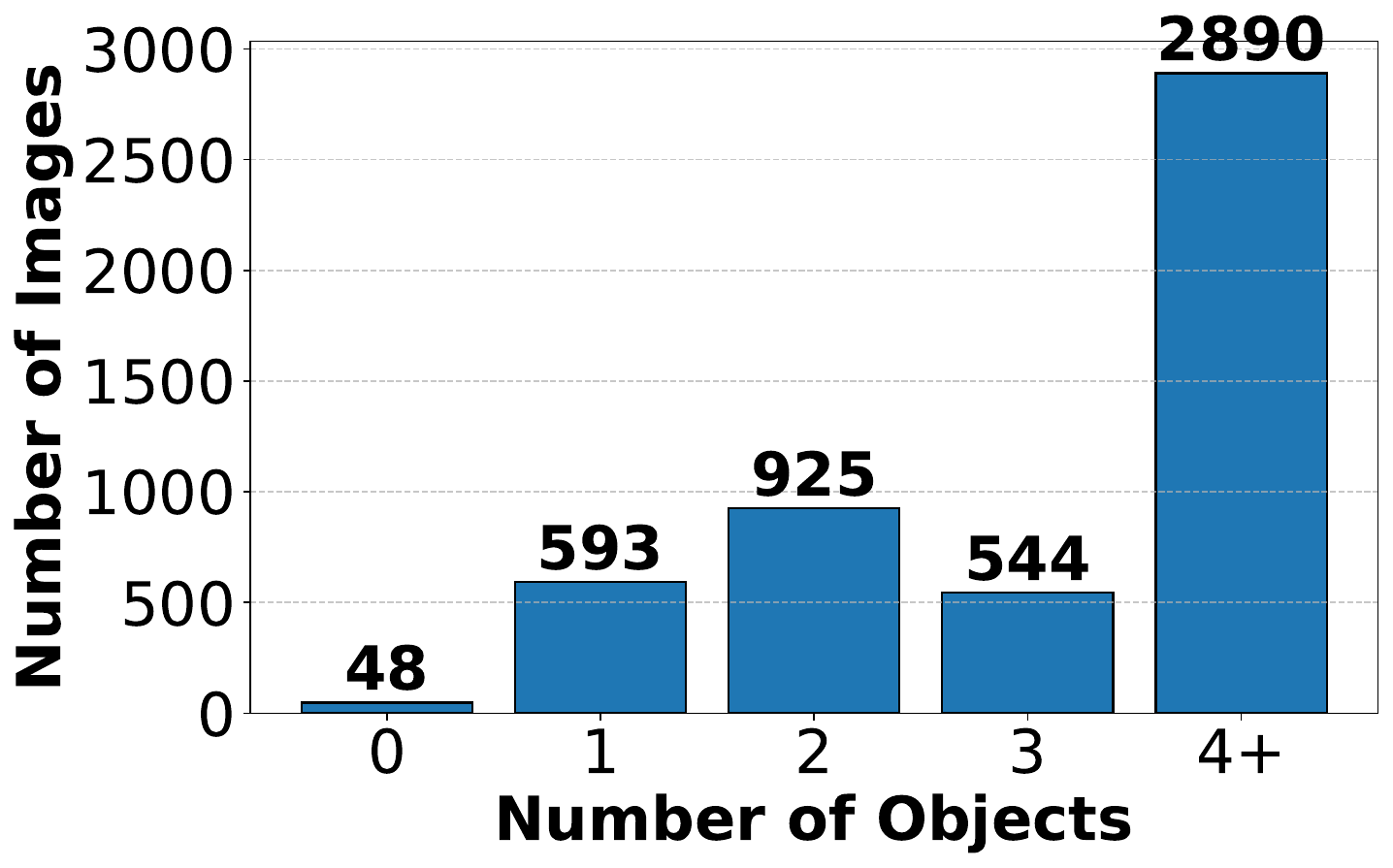}}
\caption{Distribution of Object Counts / Image in COCO dataset.}
\label{objects_number_distribution}
\end{figure}

In this study, we utilize three distinct datasets, two of which are derived from the \textit{COCO validation set}~\cite{Lin2014} and video stream frames, to evaluate the routing algorithm. The first is the \textbf{Original COCO Validation Dataset}, consisting of 5,000 images annotated with ground truth labels, including object classes and bounding boxes spanning 80 object categories. As this dataset reflects natural scene variability, the number of objects per image varies significantly. Fig.~\ref{objects_number_distribution} illustrates the distribution of object counts, highlighting this diversity.

The OB routing algorithm is designed to prioritize continuity, favoring sequences of visually similar images with minimal variation in object density. To support this principle, we constructed a secondary dataset from the COCO validation set, termed the \textbf{Balanced Sorted Dataset}. This dataset consists of 1,000 images, evenly distributed across five categories based on the number of objects per image: `0', `1', `2', `3', and `4 or more'. Each category contains 200 images selected to maintain uniform object counts within groups. Specifically, Group 1 comprises images with no objects; Group 2, exactly one object; Group 3, two objects; Group 4, three objects; and Group 5, four or more objects. When a group contains fewer than 200 unique images, random duplications are introduced to reach the target count. The images are then sent to the gateway sequentially, ordered by group.

We also evaluated our routing pipeline on a real-world \textbf{video stream}  of pedestrians crossing the road by treating it as a detection dataset. Because no manual annotations were available, we decoded the raw file and extracted every frame (at the original frame rate) into a sequence of JPEG images. We then generated the ground-truth annotations by running the powerful YOLOv8x model over each frame, saving its detections in YOLO-format to serve as our reference labels.

\subsubsection{Testbed}

\begin{figure}[t]
\centerline{\includegraphics[width=0.50\textwidth]{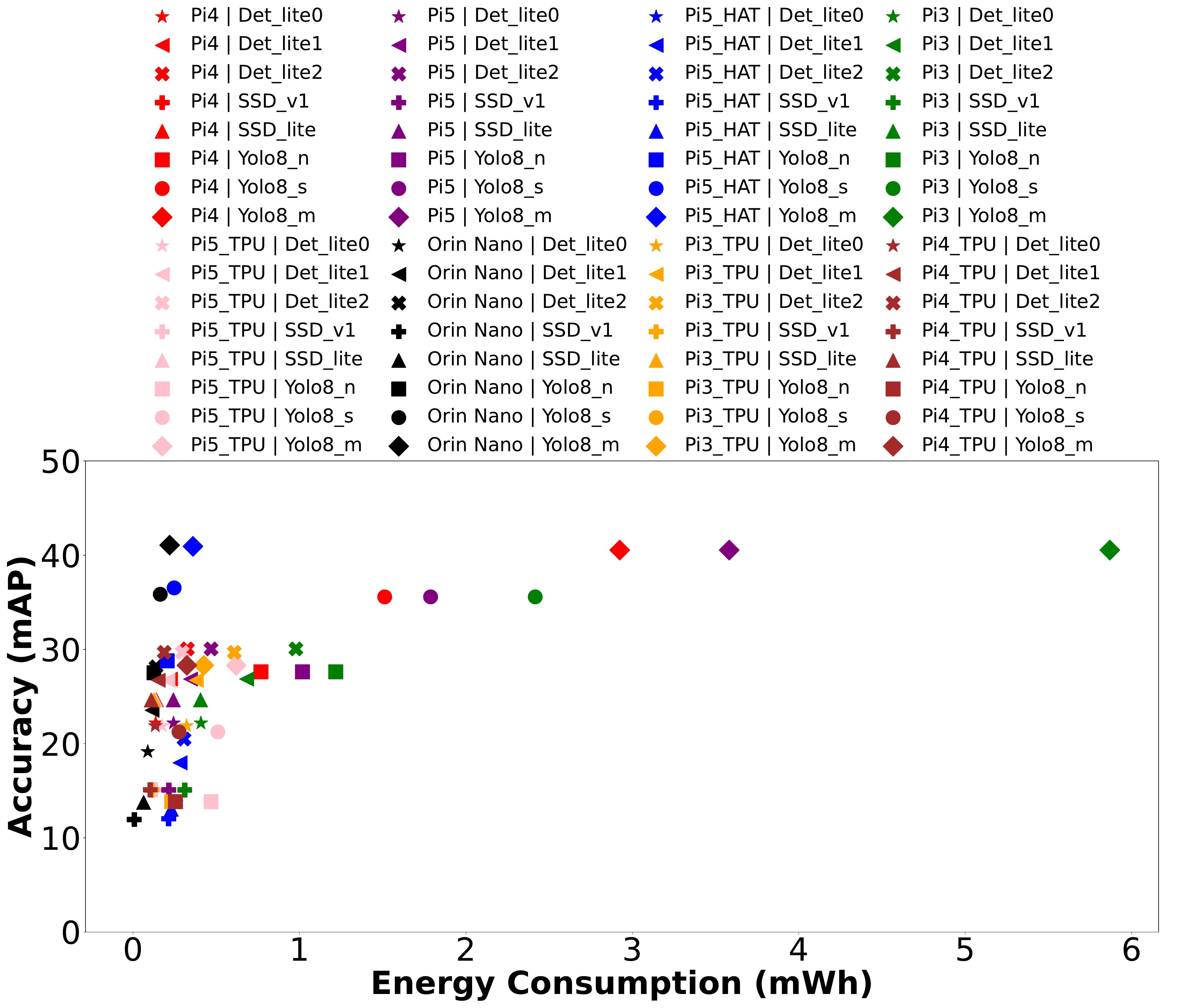}}
\caption{Pareto frontiers comparing object detection models across various edge devices.}
\label{fig:partofront}
\end{figure}

This section outlines the experimental testbed used to evaluate the proposed routing approach. The testbed configuration builds on exhaustive prior benchmarking results~\cite{daghash2024}, which involved a comprehensive and resource-intensive evaluation of various combinations of edge devices and object detection models across three key performance metrics: \textit{energy consumption}, \textit{inference time}, and \textit{mAP}. This benchmarking effort, which profiled a wide range of devices and models, provides an invaluable reference point for the research community studying edge-based object detection. Specifically, the devices evaluated included Raspberry Pi 3 (with and without TPU), Raspberry Pi 4 (with and without TPU), Raspberry Pi 5 (with and without TPU and with an AI HAT), and the Jetson Orin Nano, making a total of eight. The eight profiled object detection models spanned from lightweight variants—SSD v1, SSD Lite, EfficientDet-Lite0, Lite1, and Lite2—to the YOLOv8 family (Nano, Small, and Medium). Together, these metrics reflect the inherent trade-offs in deploying object detection models on edge hardware. Notably, the benchmarking revealed that no single device-model pair consistently outperforms others across all criteria, reinforcing the need for context-dependent model selection guided by specific performance objectives. Fig.~\ref{fig:partofront} illustrates the accuracy–energy consumption trade-offs for all 64 device-model combinations.

To construct our testbed, we selected only device-model pairs that lie on or near the Pareto front in at least one performance dimension. We repeated all the configurations in Fig.~\ref{fig:partofront} for five object groups—`0', `1', `2', `3', and `4 or more' objects per image—an effort that required extensive experimentation and significant computational time. This ensures that all included devices offer meaningful trade-offs—i.e., each is strong in at least one metric—while excluding configurations that are consistently suboptimal across all criteria. Energy efficiency and latency do not vary across groups. Therefore, for energy efficiency, we selected Jetson Orin Nano running SSD v1 which demonstrated the lowest energy consumption among all configurations. In terms of inference latency, the Raspberry Pi 5 with TPU paired with SSD v1 achieved the shortest inference time. For detection accuracy~(mAP), the optimal configuration varied across object count groups, underscoring the need for adaptive routing. Specifically, for Group 1 (images with no objects), the Raspberry Pi 5 with TPU running SSD v1 achieved the highest mAP. For Group 2 (images with one object), the Raspberry Pi 5 with TPU running SSD Lite performed best. In Group 3 (images with two objects), the Jetson Orin Nano with YOLOv8-small yielded the highest accuracy. Finally, for Groups 4 and 5 (images with three and four or more objects), the Raspberry Pi 5 equipped with an AI Hat running YOLOv8-small demonstrated superior performance. Table~\ref{tab:testbed_pairs} provides the finalized list of device-model pairs, and Table~\ref{tab:testbed_config} details the corresponding device specifications. This testbed supports a thorough assessment of the routing algorithm’s effectiveness in achieving optimal trade-offs between energy efficiency and performance across a variety of edge computing environments.

\begin{table}[htbp]
\scriptsize
\caption{Experimental Testbed Configurations}
\centering
\label{tab:testbed_pairs}
\begin{tabular}{|c|c|c|c|}
\hline
\textbf{Metrics} & \textbf{Edge Device} & \textbf{Object Detection} & \textbf{Runtime}\\
\hline
Energy Cons. & Jetson Orin Nano     & SSD v1 & TensorRT                        \\ \hline
Inference Time     & Raspberry Pi 5 + TPU      & SSD v1 &TFLite                        \\ \hline
mAP – Group 1 & Raspberry Pi 5 + TPU     & SSD v1  &TFLite                       \\ \hline
mAP – Group 2  & Raspberry Pi 5 + TPU           & SSD Lite &TFLite                       \\ \hline
mAP – Group 3 & Jetson Orin Nano         & YOLOv8-small    &TensorRT               \\ \hline
mAP – Group 4 & Raspberry Pi 5 + AI Hat  & YOLOv8-small   & HEF                \\ \hline
 mAP – Group 5 & Raspberry Pi 5 + AI Hat & YOLOv8-small     & HEF          \\ 
\hline
\end{tabular}
\vspace{-0.3cm}
\end{table}

\begin{table}[t]
\scriptsize
\centering
\caption{Testbed Device Specifications}
\label{tab:testbed_config}
\begin{tabular}{|c|c|c|c|c|c|}
\hline
\textbf{Device Name}        & \textbf{Unit}        & \textbf{Mem.}  & \textbf{OS/SDK} &\textbf{Bluetooth} &\textbf{Wi-Fi}    \\
\hline 
Pi 4 & CPU      & 4 GB   &Debian Bookworm & Off & Off     \\ \hline
Pi 5 + TPU  & TPU  & 4 GB &Debian Bookworm   & Off & Off    \\ \hline
Pi 5 + TPU  & TPU  & 4 GB &Debian Bookworm   & Off & Off         \\ \hline
Pi 5 + AI Hat     & NPU  & 4 GB &Debian Bookworm   & Off & Off       \\ \hline
Orin Nano            & GPU    & 8 GB   &JetPack 5.1.3   & Off &Off \\ \hline
Orin Nano            & GPU    & 8 GB   &JetPack 5.1.3   & Off &Off \\ 
\hline
\end{tabular}
\vspace{-0.3cm}
\end{table}

\subsection{Experimental Procedure}
To the best of our knowledge, no existing work in the literature directly addresses the same problem with a comparable approach. Therefore, 
to rigorously evaluate the effectiveness of our proposed routing strategies, we systematically compared them against a set of baseline routing methods, 
 as well as several widely used strategies: 

    \noindent \textbf{--Oracle Router (Orc)}: Serves as an ideal benchmark by assuming perfect prior knowledge of the number of objects in each incoming image. This is achieved by leveraging the ground truth annotations. The true object count is passed as metadata alongside the image request to the gateway, enabling the routing algorithm to select the optimal device-model pair by applying the \(\delta\)mAP criterion. This router reflects the upper bound of routing performance under perfect estimation.
    
  \noindent \textbf{--Round Robin Router (RR)}: Distributes incoming image requests sequentially in a round robin fashion across all available nodes, regardless of their energy efficiency or performance.
  
  \noindent\textbf{--Random Router (Rnd)}: Assigns requests to a node randomly, independent of any system metrics or workload history.
  
  \noindent \textbf{--Lowest Energy Router (LE)}: Routes each image to the device-model pair that consumes the least energy, prioritizing power efficiency over accuracy or latency.
  
  \noindent \textbf{--Lowest Inference Time Router (LI)}: Selects the node with the shortest average inference time, aiming to minimize latency without considering energy or accuracy.
  
  \noindent \textbf{--Highest mAP Router (HM)}: Always chooses the device-model pair with the highest mAP, independent of the object count in the image.
  
  \noindent \textbf{--Highest mAP per Group Router (HMG)}: Within each object-count group, the device-model pair achieving the highest mAP is selected, irrespective of energy consumption.

Given the varying performance requirements of different real-world applications, we evaluated the proposed and baseline routers across a range of \(\delta\)mAP values (e.g., 0, 5, 10, 15, 20, and 25). A value of \(\delta\)mAP = 0 enforces strict accuracy, whereas higher values allow greater flexibility in selecting less accurate but more energy-efficient models.

To ensure consistent load generation, we used Locust\footnote{https://locust.io/} to send image requests back-to-back in a  piggybacked fashion where each new request is triggered only after the previous image has been processed and a response is received, simulating continuous workloads. Performance is evaluated using energy consumption as the primary metric, with accuracy as a constraint, while latency and gateway overhead are reported as supporting quality-of-service (QoS) indicators:

    \noindent \textbf{--Energy Consumption}: Measured in milliwatt-hours (mWh) over the full duration of the experiment. To isolate the energy used for processing, we subtracted the baseline idle energy consumption of all devices when no tasks are executed but the devices remain powered on.
    
    \noindent \textbf{--Latency}: Reported in seconds as the total time required to complete all image requests from start to finish.
    
    \noindent \textbf{--Accuracy}: Evaluated using the mAP, computed via the FiftyOne tool\footnote{https://docs.voxel51.com/}, which provides robust comparisons against ground truth annotations.
    
    \noindent \textbf{--Gateway Overhead}: To assess the computational cost and overhead of different routing strategies, we separately report the energy consumption (mWh) and latency (seconds) incurred within the gateway during the routing decision-making process. This metric isolates the overhead introduced by each router, highlighting the trade-offs between routing complexity and performance.


Experiments were conducted using the three datasets described earlier: (1) the original COCO validation set, which naturally reflects varied object distributions; (2) the balanced and sorted dataset was constructed to include an equal number of images for each object count group, with images grouped by object count—i.e., all images containing zero object first, followed by those with one object, and so on—thus collectively favoring the output-based approach; and (3) a pedestrian-crossing video dataset, where every frame of a real-world road-crossing sequence was extracted as an image. This triple-dataset evaluation enables a comprehensive analysis of routing behaviour under both real-world and controlled conditions.

\subsection{Experimental Results}

\subsubsection{Full COCO Dataset Results:}

\begin{figure*}[htbp]
    \centering
    \subfigure[Accuracy (mAP)]{
        \includegraphics[width=0.30\textwidth]{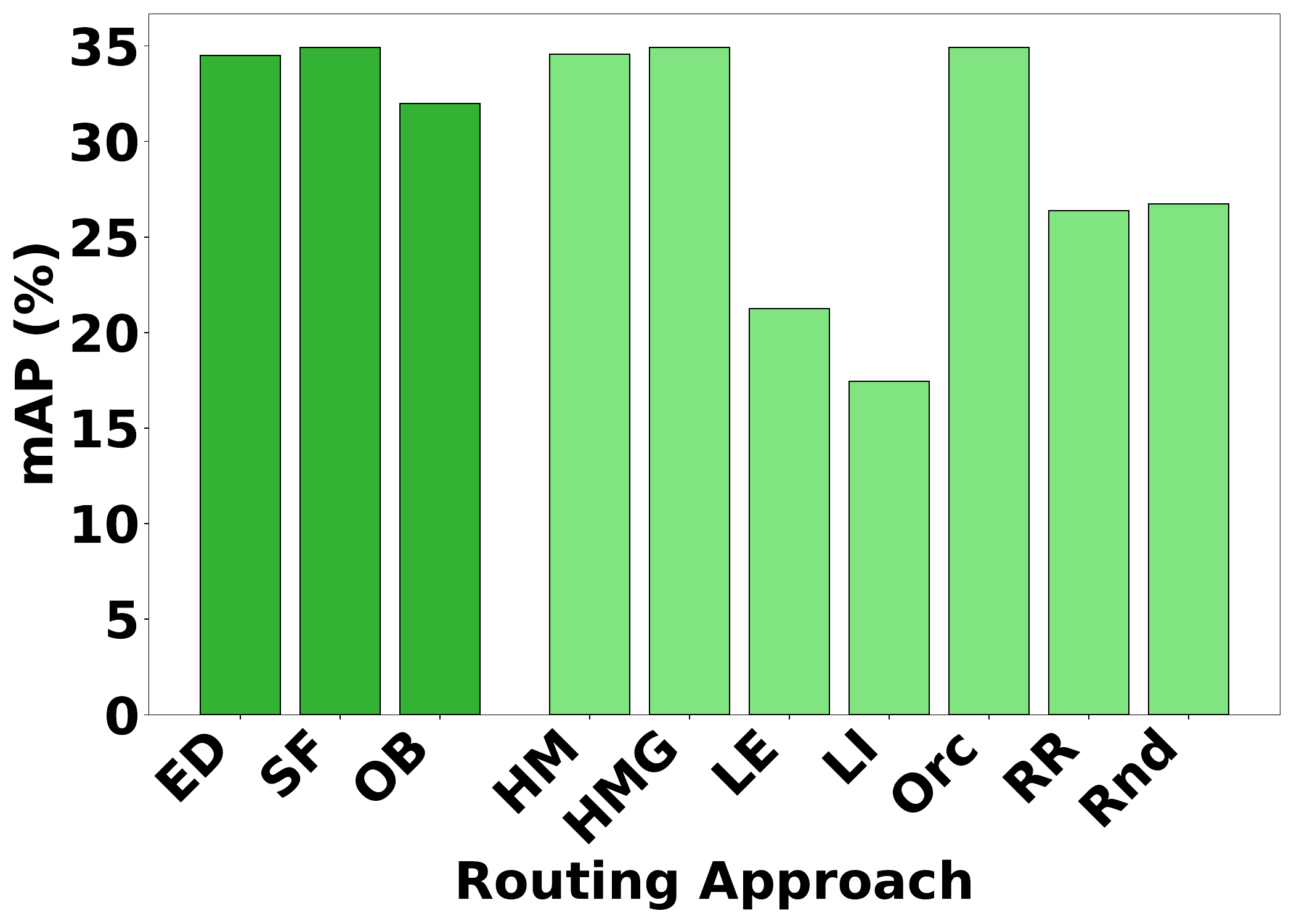}
    }
    \subfigure[Energy Consumption]{
        \includegraphics[width=0.30\textwidth]{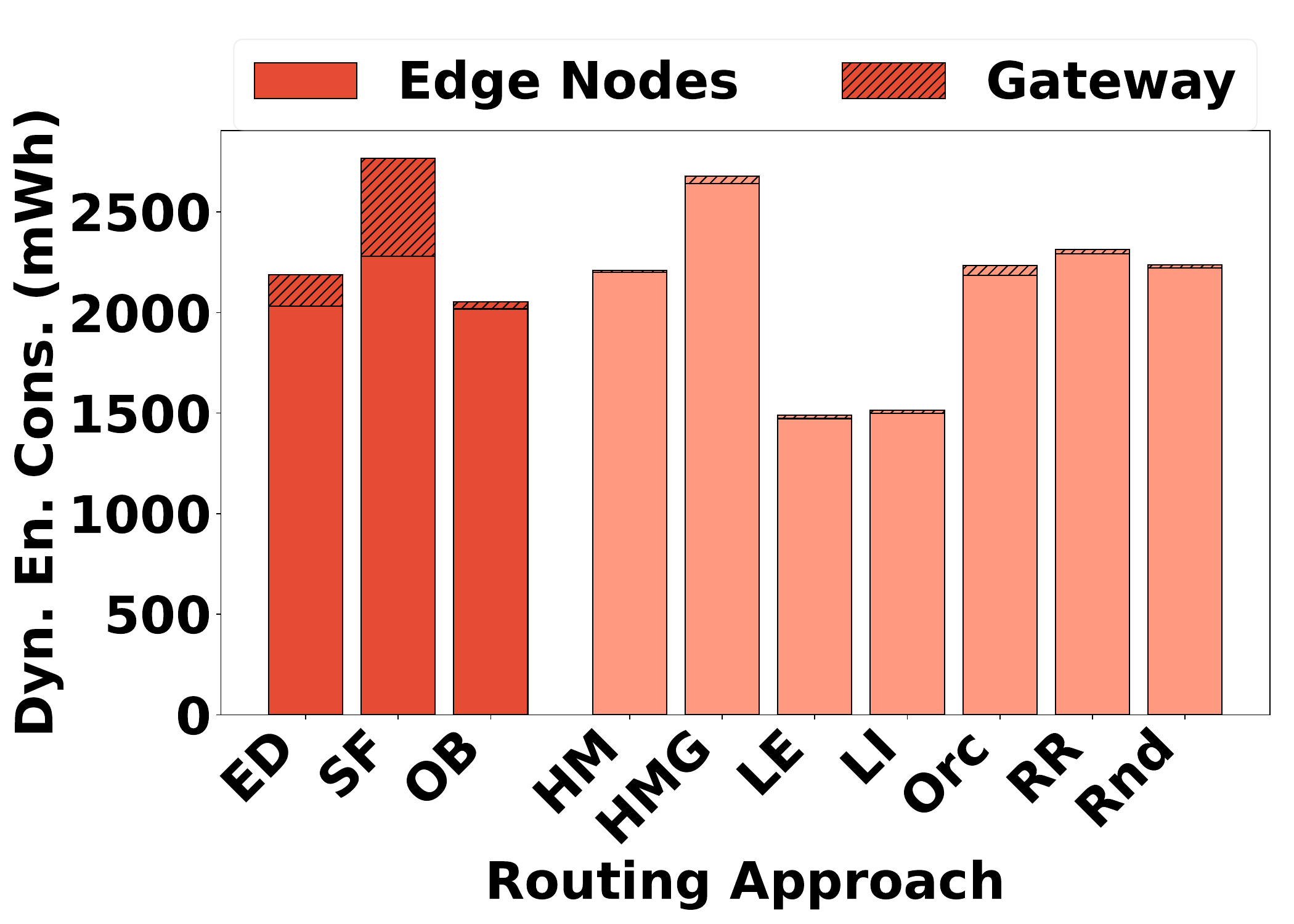}
    }
    \subfigure[Latency]{
        \includegraphics[width=0.30\textwidth]{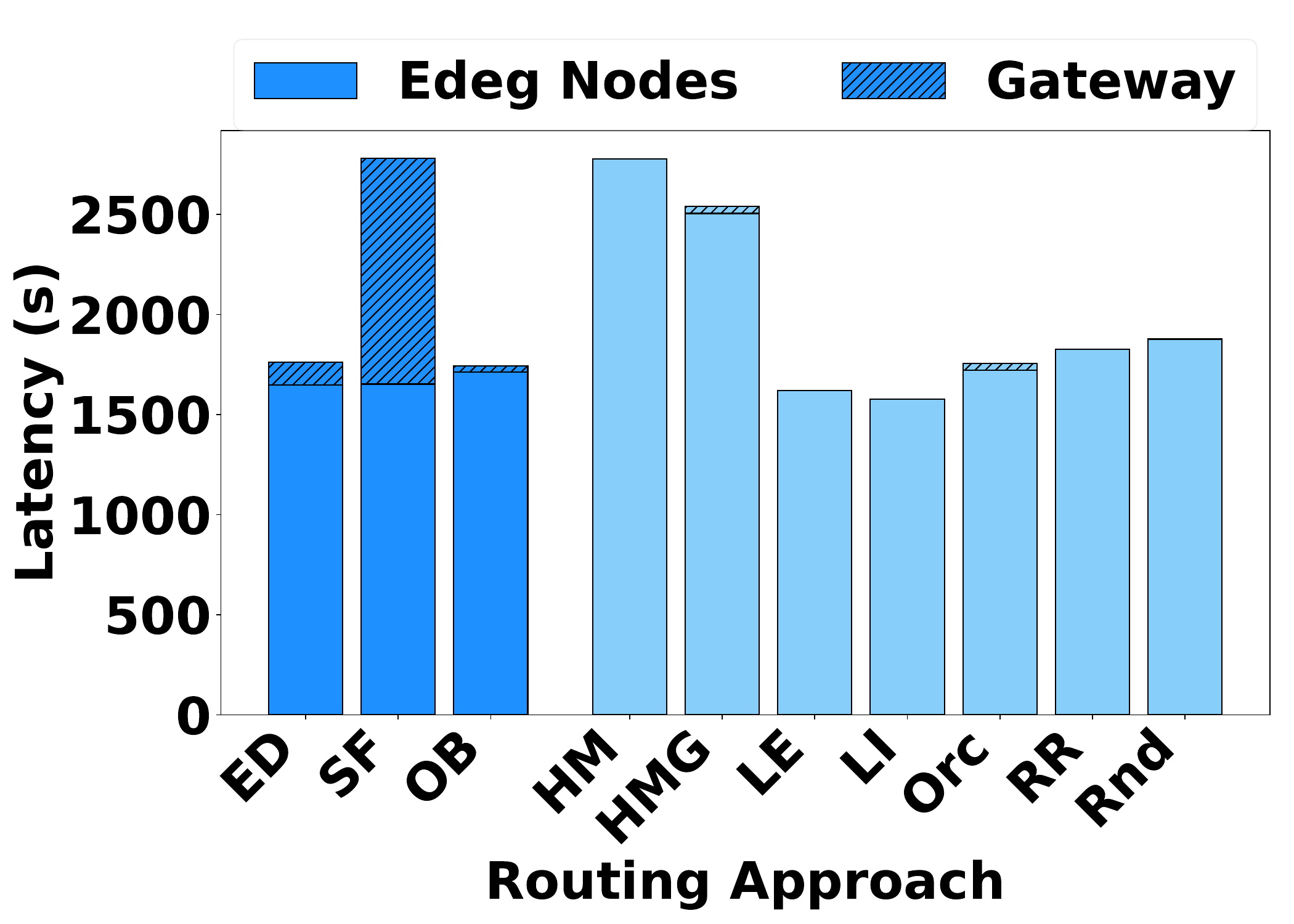}
    }

    \caption{Accuracy (mAP), Latency and Dynamic Energy Consumption for proposed routing approaches and all other baselines using full coco validation dataset. (Orc=Oracle, RR=Round Robin, Rnd=Random, LE=Lowest Energy, LI=Lowest Inference, HM=Highest mAP without considering Groups, HMG=Highest mAP Per Group, ED=Edge Detection, SF=SSD-Based Front, OB=Output-Based) at \(\delta\)mAP=5.}
    \label{fig:results_full_coco_dataset}
    \vspace{-0.2cm}
\end{figure*}

To evaluate the performance of the proposed routing approaches in comparison to baseline methods, we conducted experiments using the full COCO validation dataset, with the \(\delta\) mAP threshold set to 5 unless otherwise specified. The results across energy consumption, latency, and detection accuracy are presented in Fig.~\ref{fig:results_full_coco_dataset}. In terms of dynamic energy consumption Fig.~\ref{fig:results_full_coco_dataset}(b), LE achieves the lowest energy usage, establishing the theoretical lower bound. LI follows closely, consuming only marginally more. On the other end of the spectrum, HMG, achieving best accuracy per group, increases energy consumption by approximately 80 percent over the lower bound.  
The proposed SF exhibits an even greater rise of nearly 85\%, attributed to the additional overhead introduced by gateway analysis. The Orc router results in only 50\% increase in energy usage. Among the proposed methods, ED and OB increase consumption by about 45\% and 37\%, respectively. In comparison, RR and Rnd lead to energy increases of around 55\% and 50\%.

With respect to detection accuracy (mAP), Fig.~\ref{fig:results_full_coco_dataset}(a) shows that HMG achieves the highest score, setting the upper bound. Orc and SF perform similarly, with accuracy dropping by less than one percent. HM and ED experience slight reductions of approximately 1\% and 2\%, respectively. A more significant 9\% drop is observed in the OB router’s performance, attributed to the highly variable and unpredictable object counts present in the COCO dataset. Finally, RR and Rnd incur significant drops of around 25\%, while LE and LI show the largest losses—about 40\% and 50\%, respectively. Results show that the proposed routing approaches, such as ED and OB, can achieve mAP values very close to optimal methods like HMG, while also delivering significant energy savings of approximately 45\% and 37\%, respectively. 

Latency results, measured as the total time to process the entire dataset, are presented in Fig.~\ref{fig:results_full_coco_dataset}(c). As expected, LI achieves the lowest overall latency, with LE following closely—incurring an increase of less than 1\%. In contrast, HM and SF exhibit the highest latency increases, each rising by approximately 75\%, while HMG incurs a latency overhead of around 60\%.. In contrast, Orc, OB, and ED introduce relatively small latency increases of about 11\%.

\begin{insight}
    Considering upper bound on achievable mAP (as given by HMG) and the minimum energy consumption (as achieved by LE), the proposed ED approach offers the most efficient trade-off on static-image datasets such as COCO. It attains accuracy close to the Orc performance-which has full knowledge of object counts-while maintaining competitive energy efficiency and incurring only minimal latency overhead.
\end{insight}

\subsubsection{Balanced Sorted Dataset Results}
\begin{figure*}[htbp]
    \centering
    \subfigure[Accuracy (mAP)]{
        \includegraphics[width=0.31\textwidth]{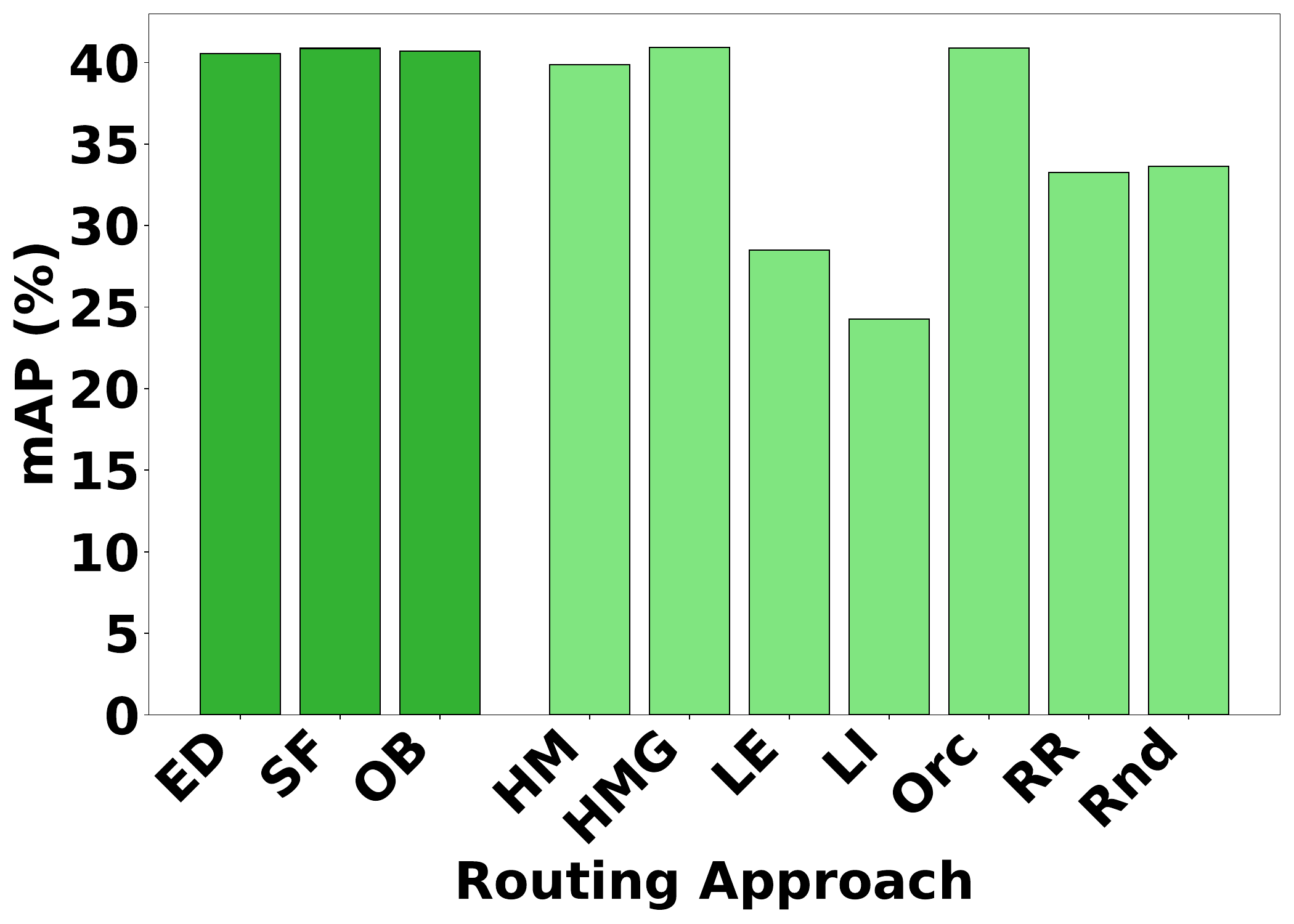}
    }
    \subfigure[Energy Consumption]{
        \includegraphics[width=0.31\textwidth]{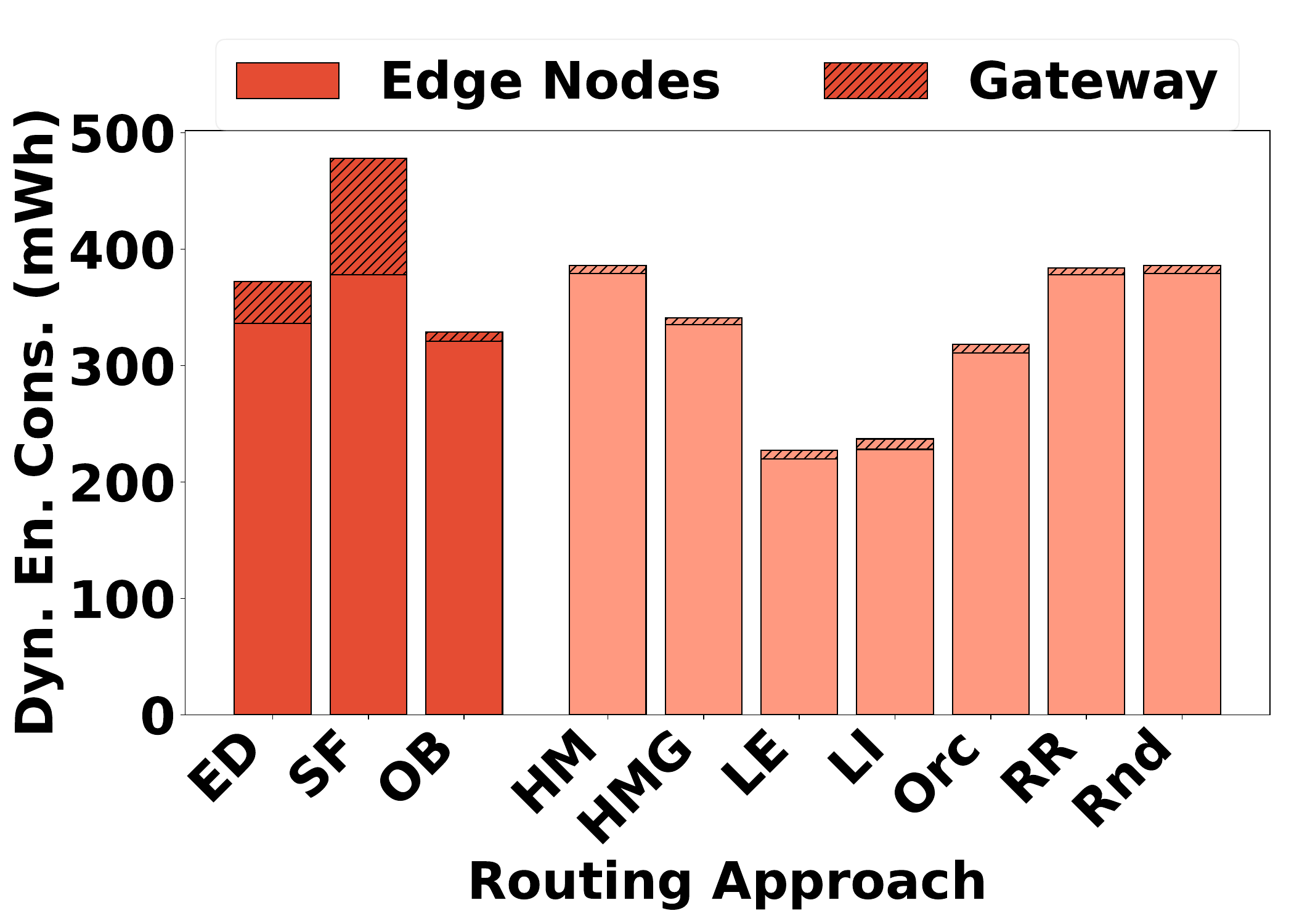}
    }
    \subfigure[Latency]{
        \includegraphics[width=0.31\textwidth]{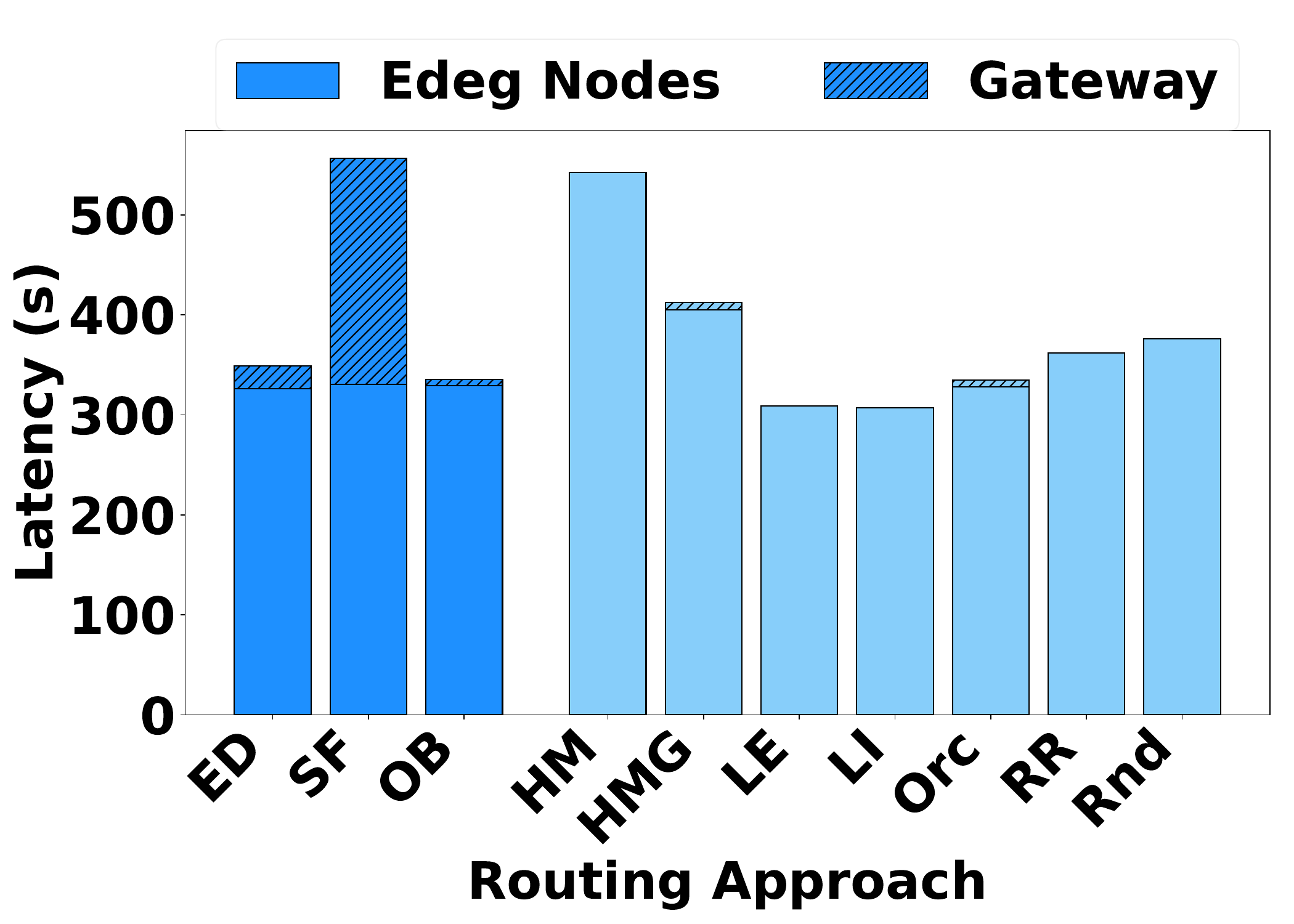}
    }

    \caption{Accuracy, Latency and Energy of the proposed and baselines approaches using sorted balanced dataset at \(\delta\)mAP=5.}
    \label{fig:results_sorted_balanced_dataset}
\end{figure*}

This section highlights the results of evaluating our proposed routing approaches and baseline routing strategies using the balanced sorted dataset as shown in Fig.~\ref{fig:results_sorted_balanced_dataset}. With respect to dynamic energy consumption Fig.~\ref{fig:results_sorted_balanced_dataset}(b), LE demonstrates the most energy-efficient performance, consuming only 227~mWh and establishing the theoretical lower bound. LI performs similarly, with only a slight increase in consumption. At the opposite end, HMG exhibits a substantial rise in energy usage—approximately 50\% higher than the baseline—while SF records more than double the energy consumption of LE. The Orc router shows a 40\% increase in energy use relative to the lower bound. Among the proposed approaches, ED and OB consume 64\% and 45\% more energy, respectively. In comparison, the baseline methods RR and Rnd result in increases of roughly 70\% and 62\%.

Latency results, presented in Fig.~\ref{fig:results_sorted_balanced_dataset}(c), show that LI achieves the lowest total latency at approximately 306 seconds, with LE following closely behind, incurring only a 2-second increase. HM and SF exhibit the highest latency overheads, with increases of around 77\% and 81\%, respectively. HMG also shows a notable rise, adding approximately 34\% to the baseline latency. In contrast, Orc and OB introduce modest increases of about 9\%, while ED shows a slightly higher increase of 13\%. The RR and Rnd routers fall in the mid-range, with latency increases of 18\% and 22\%, respectively.

In terms of detection accuracy (mAP), Fig.~\ref{fig:results_sorted_balanced_dataset}(a) indicates that HMG achieves the highest score at 40.94, establishing the upper performance bound. Orc, SF, and OB closely follow, each exhibiting a negligible decline of less than 1\%. ED also shows minor reductions of approximately 1\%. In contrast, RR and Rnd suffer notable accuracy losses of around 18\%, while LE and LI demonstrate the most substantial declines, with reductions of approximately 30\% and 40\%, respectively.

\begin{insight}
  Among the proposed routing methods, OB offers the best overall trade-off, closely approaching oracle-level performance in both accuracy and latency while achieving the lowest energy consumption of all proposed approaches. This makes OB the most suitable choice for sequential image processing scenarios.
\end{insight}

\subsubsection{Video Dataset Results}
\begin{figure*}[htbp]
    \centering
    \subfigure[Accuracy (mAP)]{
        \includegraphics[width=0.31\textwidth]{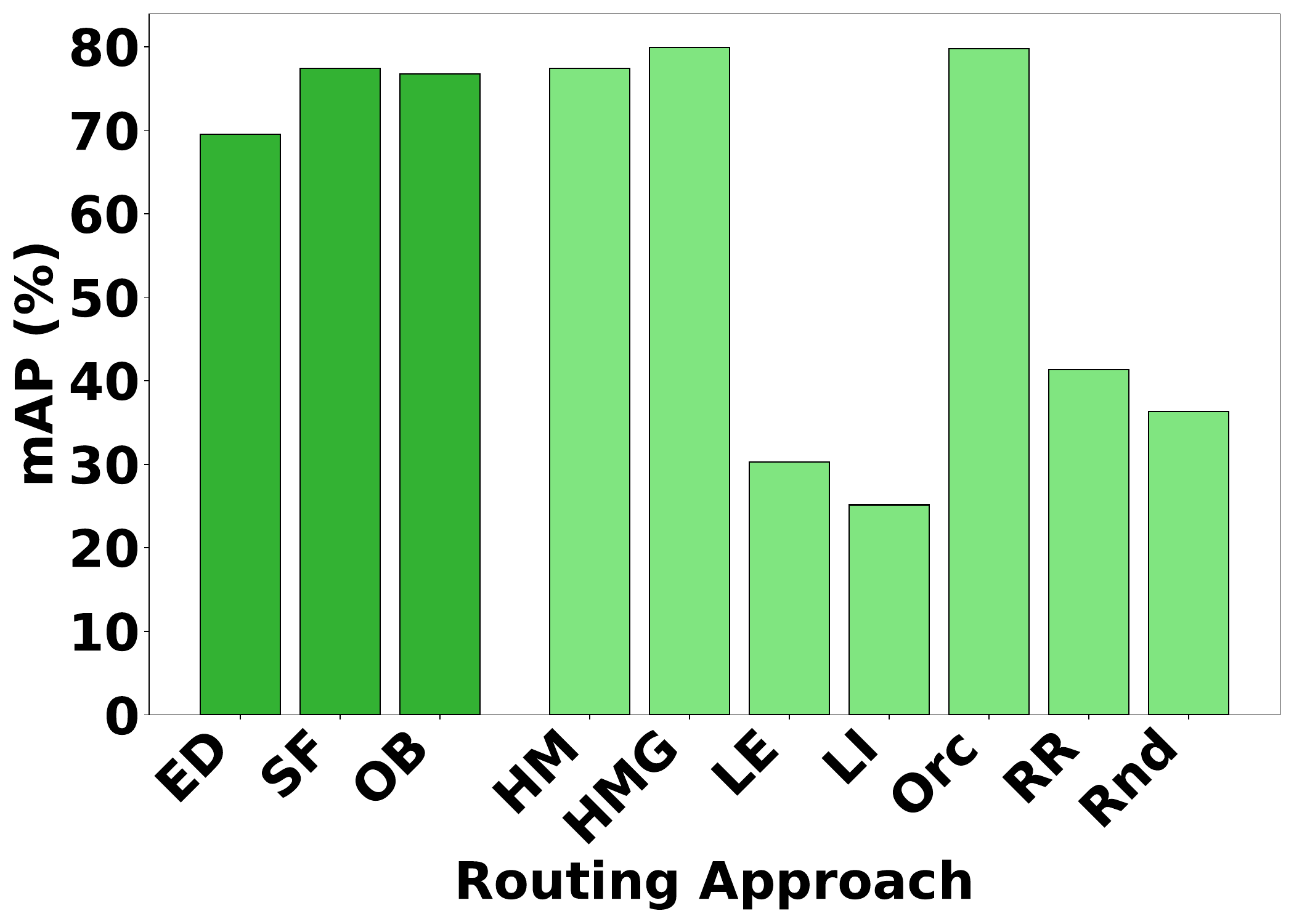}
    }
    \subfigure[Energy Consumption]{
        \includegraphics[width=0.31\textwidth]{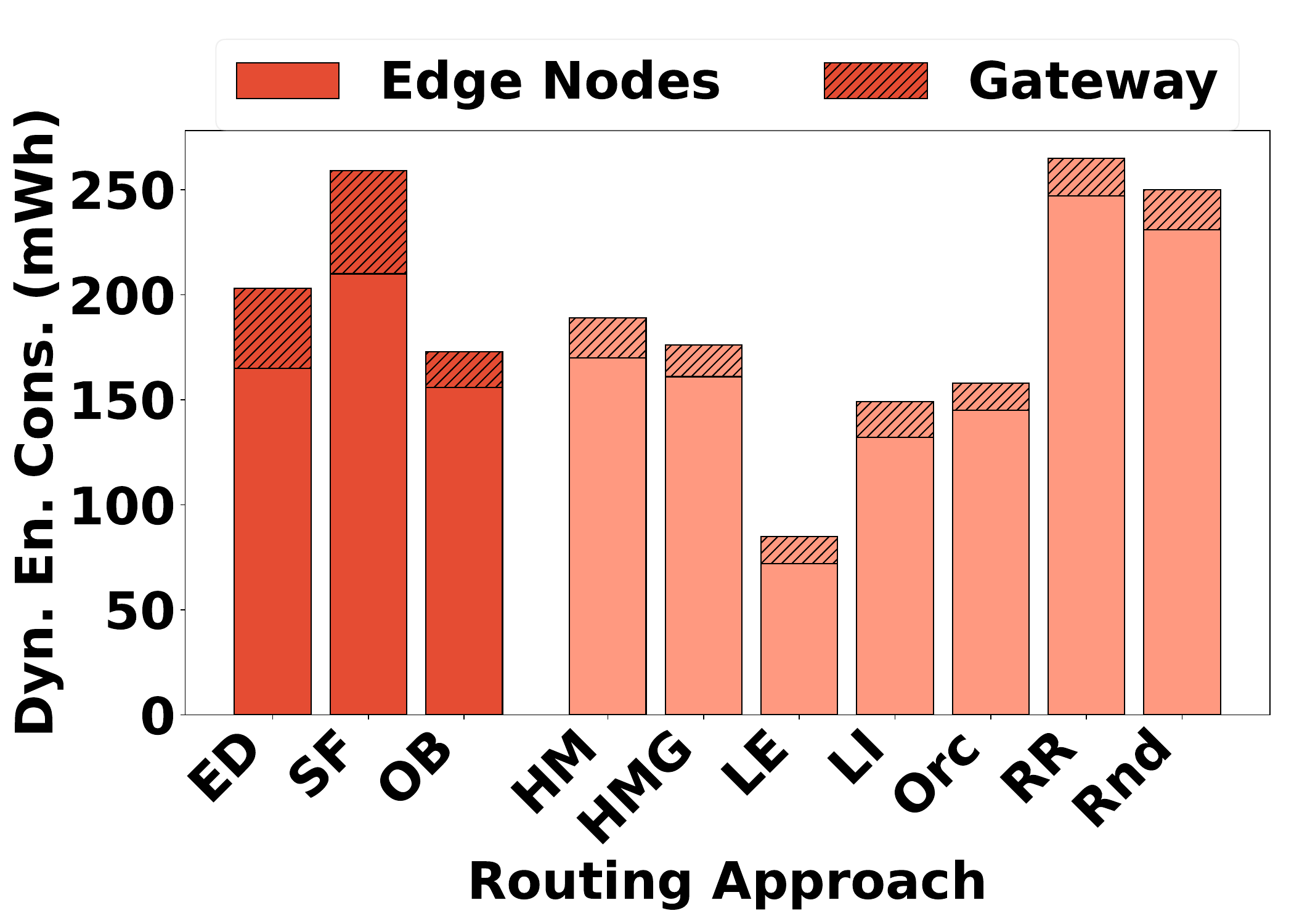}
    }
    \subfigure[Latency]{
        \includegraphics[width=0.31\textwidth]{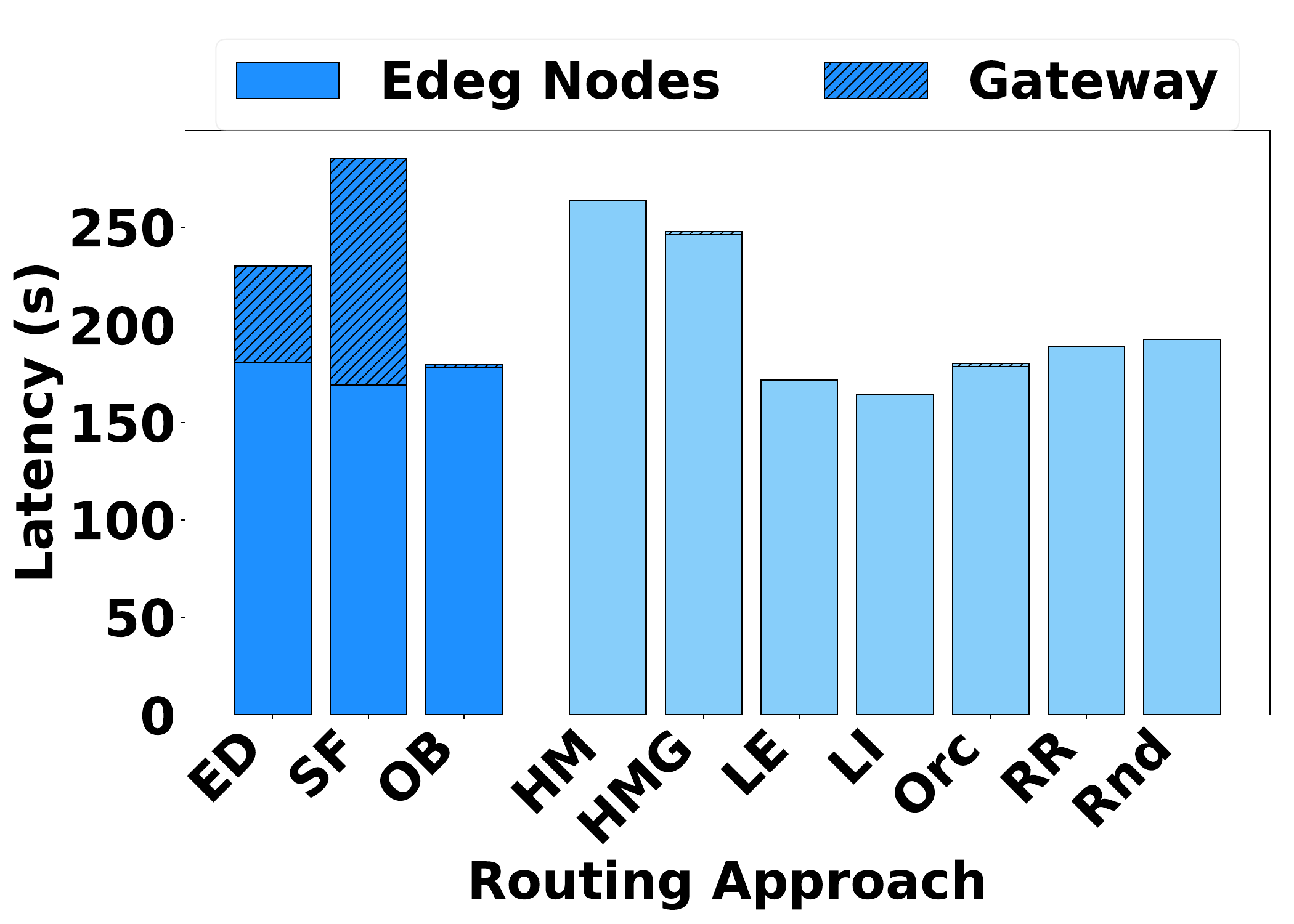}
    }

    \caption{Accuracy, Latency and Energy of the proposed and baselines approaches using video dataset at \(\delta\)mAP=5.}
    \label{fig:results_video_dataset}
    \vspace{-0.2cm}
\end{figure*}
This section presents the results of the different routing approaches evaluated using the pedestrian video dataset. The corresponding Fig.~\ref{fig:results_video_dataset} illustrates the performance trends across key metrics. With respect to dynamic energy consumption Fig.~\ref{fig:results_video_dataset}(b), LE consumes only 85~mWh. LI follows with a 75\% increase in consumption over LE. At the opposite end, HMG shows a substantial rise in energy usage—approximately 100\% higher than the baseline—while SF consumes more than three times the energy of LE, exceeding 200\%. The Orc router results in an 85\% increase relative to the lower bound. Among the proposed methods, ED and OB increase energy usage by approximately 138\% and 100\%, respectively. For comparison, the baseline methods RR and Rnd show similar trends, each exceeding 200\% of the energy consumption of LE.

Latency results, illustrated in Fig.~\ref{fig:results_video_dataset}(c), indicate that LI achieves the best performance, with the lowest total latency of approximately 164 seconds. LE follows closely, incurring a minimal increase of 7 seconds. On the higher end, HM and SF experience the greatest latency overheads, with increases of approximately 61\% and 73\%, respectively. HMG also shows a significant rise, adding around 50\% over the baseline. In comparison, Orc and OB introduce only modest latency increases of about 9\%, while ED results in a more pronounced increase of 40\%. The RR and Rnd routers occupy a middle ground, with latency overheads of 29\% and 17\%, respectively.

Regrading accuracy (mAP), Fig.~\ref{fig:results_video_dataset}(a) shows that HMG achieves the highest score at 79.95, setting the upper bound for performance. Orc closely follows, with a minimal decline of less than 1\%. SF, OB, and HM each experience moderate reductions of around 4\%. ED shows a more significant drop, with accuracy decreasing by approximately 14\%. In contrast, RR and Rnd incur substantial losses of 50\% and 55\%, respectively, while LE and LI show the most severe accuracy degradation, with mAP reductions of about 63\% and 75\%.

\begin{insight}
Among the proposed methods, OB emerges as the most effective for video scenarios, combining high accuracy with low latency and reasonable energy consumption—benefiting from the continuity and temporal consistency of video frames.
\end{insight}

\subsubsection{Results Across Different 
\(\delta\)mAP Values}

\begin{figure*}[htbp]
    \centering
    \subfigure[Accuracy (mAP)]{
        \includegraphics[width=0.31\textwidth]{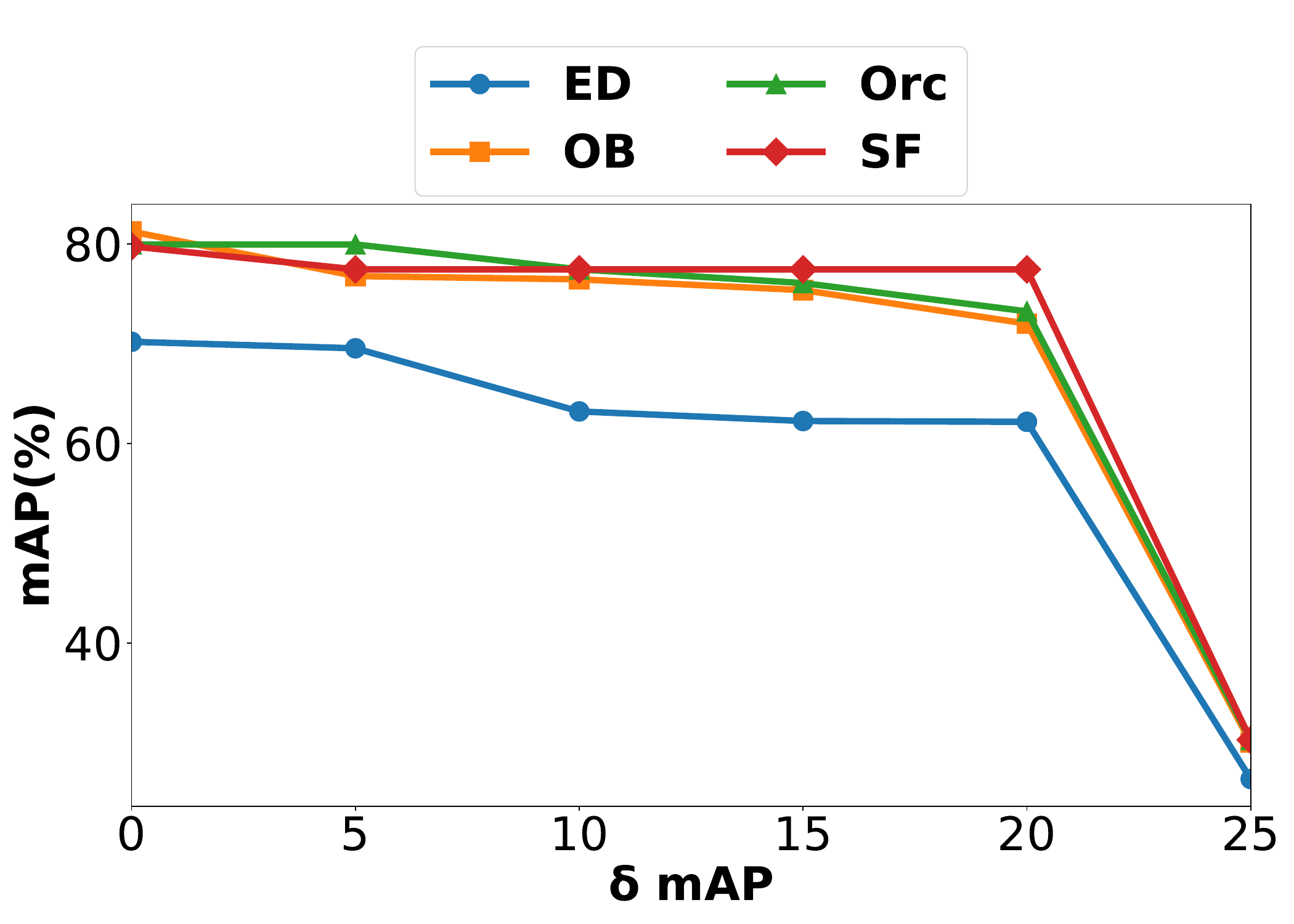}
    }
    \subfigure[Energy Consumption]{
        \includegraphics[width=0.31\textwidth]{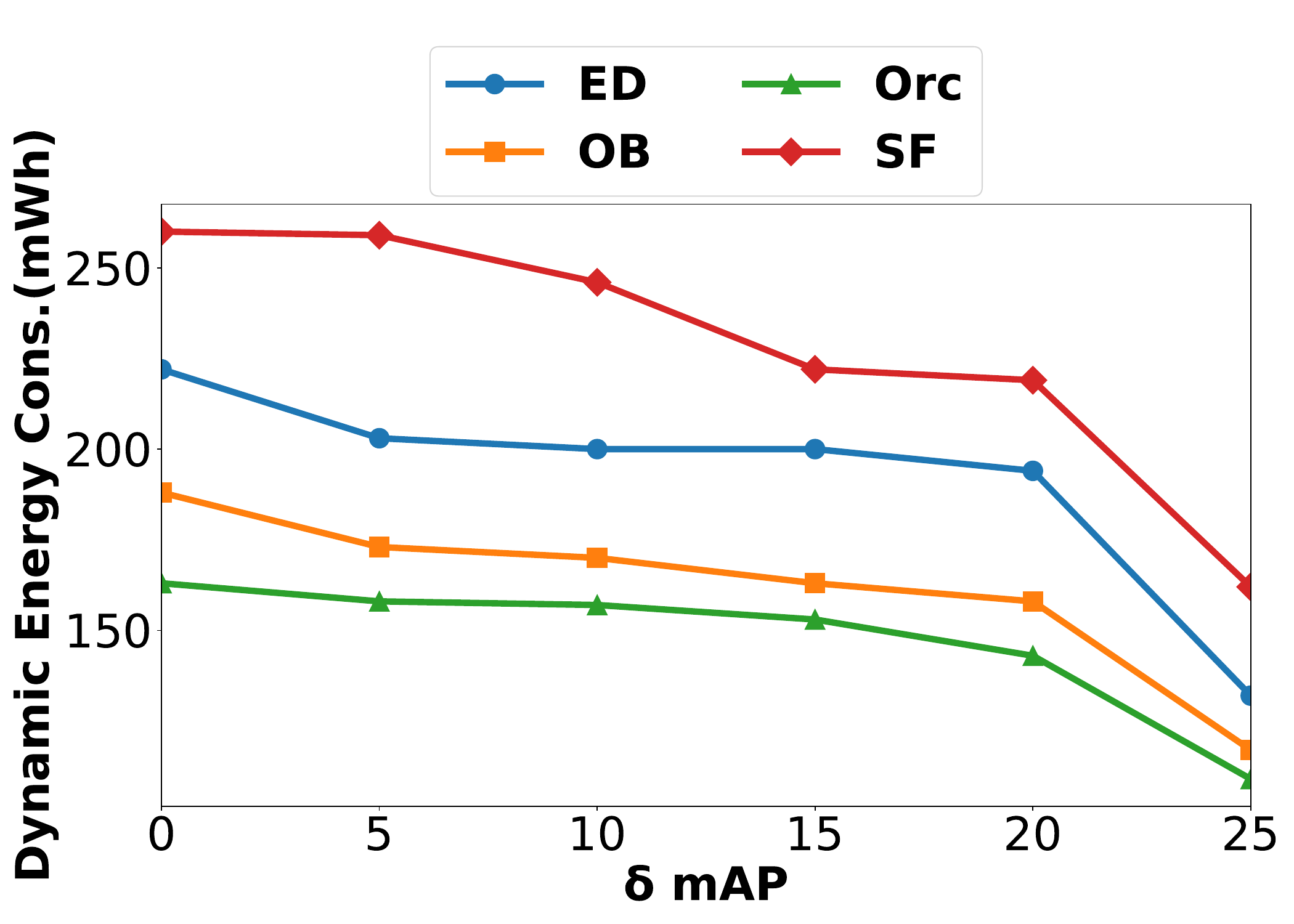}
    }
    \subfigure[Latency]{
        \includegraphics[width=0.31\textwidth]{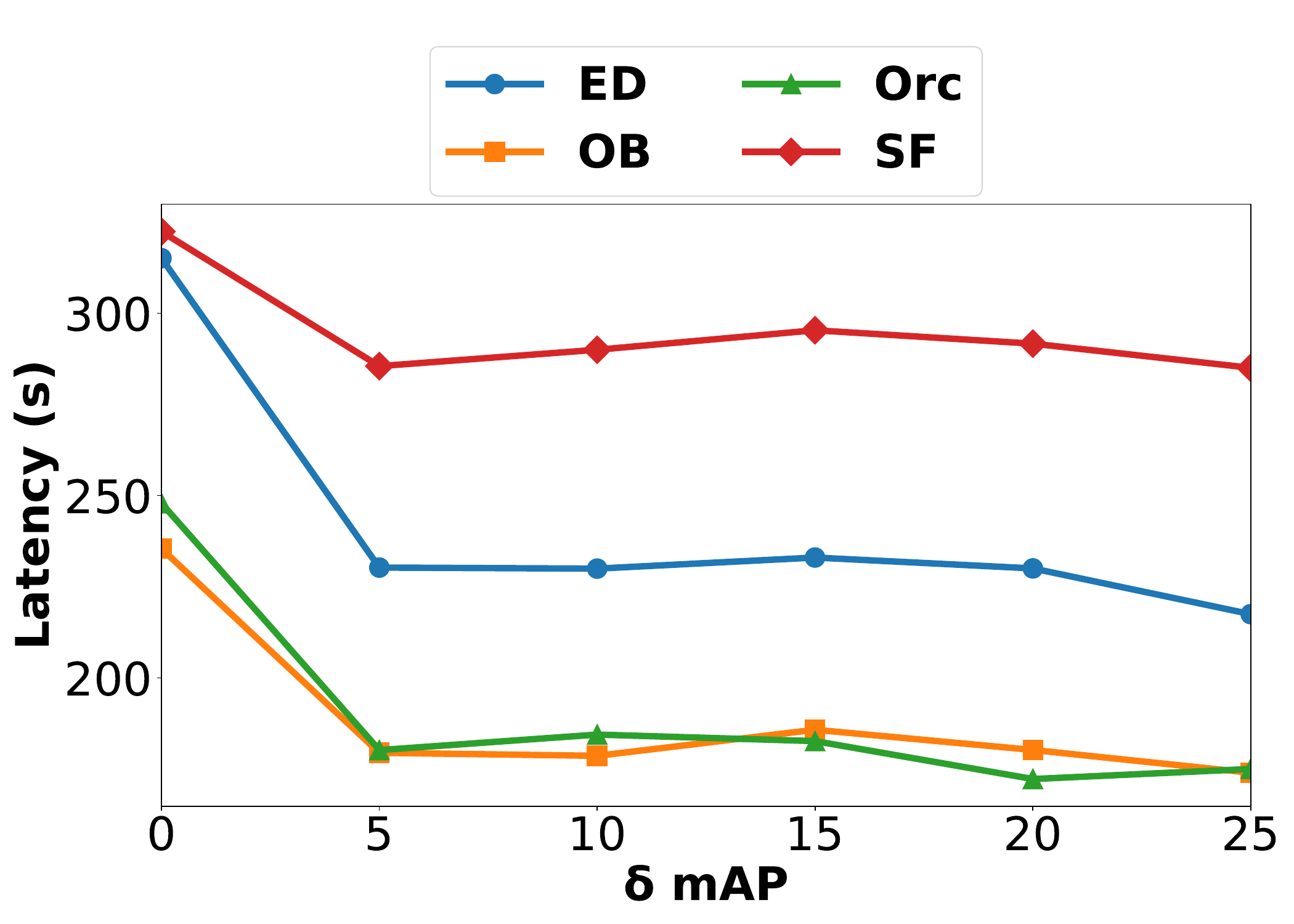}
    }

    \caption{Accuracy, Latency and Energy of the Oracle router and proposed routers for across different delta.}
    \label{fig:results_differnt_delta}
    \vspace{-0.2cm}
\end{figure*}

This section presents the performance of the Oracle and proposed routers across varying \(\delta\)mAP values. All routers including the Oracle show a substantial reduction in energy consumption when \(\delta\)mAP increases from 0 to 5. Between \(\delta\)mAP values of 5 and 20, energy consumption continues to decrease slightly, followed by a  noticeable drop at \(\delta = 25\) as Fig.~\ref{fig:results_differnt_delta} depicts.

A similar trend is observed for total latency. All routers exhibit a sharp decline in latency when \(\delta\)mAP increases from 0 to 5, with only marginal improvements observed between values of 5 and 25.

In terms of detection accuracy (mAP), most routers maintain stable performance as \(\delta\)mAP increases from 0 to 5. However, a slight decrease in mAP is observed between \(\delta = 5\) and \(\delta = 15\), followed by a significant decline between \(\delta = 20\) and \(\delta = 25\).

\begin{insight}
Results show that even setting \(\delta\)mAP to a small value such as 5,allowing up to a 5\% tolerance in accuracy drop, yields significant reductions in both energy consumption and latency, while maintaining detection performance comparable to the strictest setting (\(\delta = 0\)). Notably, with \(\delta\)mAP set to 5, the actual accuracy drop remains relatively low, around 2\%.
\end{insight} 

\subsection{Discussions}

\toolname demonstrates promising results by reducing energy consumption and latency while maintaining strong accuracy. While the Orc router represents an idealized baseline, benefiting from prior knowledge of object counts, achieving comparably strong results in practice is not infeasible. Our proposed routing approaches ED and OB achieve performance close to Oracle, with marginal differences. However, the effectiveness of each method is dataset-dependent: OB performs optimally in sequential and structured scenarios, such as the sorted balanced and video datasets, while ED is better suited for static image inputs without any continuity. 

The current design assumes static profiling, where performance metrics remain stable at runtime, which may not reflect real-world dynamics such as temperature, background load, or battery state. To handle such dynamics, profiling can be updated online by periodically collecting recent measurements and refreshing the profiles, allowing ECORE to adapt to thermal throttling and background load. Additionally, our heuristic operates at single-request granularity, limiting its applicability in batch or load-balancing contexts. It is important to note that the proposed approaches do not support multi-objective optimization; balancing trade-offs between energy and latency would require a Pareto-optimal or weighted strategy, where a greedy algorithm may no longer suffice.

\section{Related Work}
\label{sec:related work}

This section surveys related studies and situates our approach. We organize prior work into three scopes: (i) dynamic on-device adaptation, (ii) system-level orchestration across edge clusters and between edge and cloud, and (iii) routing policies. The first includes pipeline reconfiguration, local model and accelerator selection, and contextual-bandit policies within a single device. The second covers multi-stream configuration, bandwidth allocation, cluster scheduling, and offloading across edge and cloud. The third reviews query-aware dispatch among model pools in the cloud, which is conceptually related but outside our edge-vision scope.

\subsection{Dynamic on-device adaptation}

Local methods adapt inference on a single device. Pipeline and content-adaptive systems such as Chameleon~\cite{jiang2018chameleon} and NoScope~\cite{kang2017noscope} reconfigure analytics, use cascades, or apply difference detectors to reduce cost while maintaining accuracy, all within one host. Self-adaptive switching (EdgeMLBalancer~\cite{matathammal2025}, SWITCH~\cite{Marda2024}, AdaMLS~\cite{Kulkarni2023}), energy-aware design (Tundo et al.~\cite{Tundo2023}), and intra-chip selection among detectors and accelerators (SHIFT~\cite{davis2024context}) likewise operate within one device. On the other hand, these works optimize within a single device or processing pipeline rather than across heterogeneous edge devices.

\subsection{System-level orchestration}

A second line of research focuses on coordinating resources across different nodes and tiers. EdgeLeague~\cite{Tu2023} and JCAB~\cite{Zhang2022} adapt detector configurations and uplink bandwidth for multi-stream analytics under fluctuating network conditions, scheduling resources to meet QoS and latency requirements. VideoStorm~\cite{zhang2017live} performs cluster-scale scheduling based on resource and quality profiles, emphasizing throughput and deadline compliance rather than per-image content decisions. In cloud–edge offloading, CELTC~\cite{Ji2025} coordinates large cloud models with lightweight edge models using deep reinforcement learning, while Trinh et al.~\cite{Trinh2018} determine when to offload vision tasks under energy constraints. These orchestration approaches optimize configurations, bandwidth, and scheduling across the system, whereas our work operates entirely at the edge and makes content-aware selections of the most suitable device–model pair for each image.

\subsection{Routing policies} 

Outside edge vision, several studies in cloud computing explore routing among large language models using techniques such as expert dispatch, subset selection, token-level collaboration, weak supervision routing, and reinforcement learning policies~\cite{stripelis2024tensoropera, maurya2024selectllm, zheng2025citer, NEURIPS2024_e6b57a99, mohammadshahi2024routoo, sikeridis2025pickllm}. These works share the routing concept but focus on cloud-based NLP workloads, aiming to balance quality, cost, and latency, which differs from the energy-constrained and heterogeneous nature of edge vision.

Across these threads, most prior work either depends on the cloud or optimizes configuration, bandwidth, and scheduling at the system level; adapts within a single device or pipeline; or studies routing in cloud model pools outside edge vision. In contrast, \toolname operates across edge nodes and performs content-aware routing for each image to a heterogeneous device and model pair using lightweight estimators and reducing energy while preserving detection accuracy.

\section{Conclusions and Future Direction}
\label{sec:conclusions}


Deploying object detection models on resource-constrained edge devices, especially those powered by batteries or renewable energy sources such as solar panels, demands careful consideration of energy consumption. In many real-world applications, particularly safety-critical systems, the environment is highly dynamic and the number of objects captured in each frame can vary greatly. This variability directly impacts the accuracy of object detection models and highlights the need for adaptive strategies that maintain a balance between efficiency and performance. To address these challenges, we introduce \toolname  together with three dynamic, estimation-based routing algorithms. Each algorithm begins by applying lightweight preprocessing to estimate the number of objects in an incoming image request. Based on this estimate, the request is then routed to the most suitable edge device-model pair, a decision informed by detailed profiling of accuracy and energy trade-offs. This design aims to minimize overall energy consumption while ensuring that detection accuracy remains within an acceptable range. Through extensive experimental evaluation, we show that the proposed strategy consistently outperforms baseline approaches. In particular, it achieves significant reductions in energy usage without compromising accuracy, thereby demonstrating the effectiveness of dynamic, context-aware routing as a promising direction for enabling energy-efficient AI workloads at the edge.

\textbf{Future Work:} To enhance the robustness of \toolname, future work will extend the routing strategy to support batch-level decision-making for better load balancing. Additionally, incorporating multi-objective optimization techniques, such as Pareto-based or weighted approaches, will allow more flexible trade-offs between energy consumption and latency across diverse deployment scenarios.

\section*{Acknowledgment}

We would  like to acknowledge support through the Australian Research Council’s funded projects DP230100081 and LP210200213.

\bibliographystyle{ieeetr}
\bibliography{routing_paper_ref}

\end{document}